\documentclass[pra,reprint,amsmath,amssymb,showpacs,twocolumn]{revtex4-1}
\usepackage[dvipdfmx]{graphicx}
\usepackage{hyperref}
\hypersetup{
	setpagesize=false,
	bookmarksnumbered=true,%
	bookmarksopen=true,%
	colorlinks=true,%
	linkcolor=magenta,
	citecolor=magenta,
}
\usepackage{bm}
\usepackage{mathrsfs}
\usepackage{color}
\usepackage{here}
\usepackage{braket}
\usepackage{booktabs}
\usepackage{comment}
\usepackage{cases}
\usepackage{float}

\newcommand{\kk}{\bm{k}}

\newcommand{\ur}{u^{\rm{R}}}
\newcommand{\ul}{u^{\rm{L}}}
\newcommand{\tGamma}{\tilde{\Gamma}}
\newcommand{\Uc}{U_{\mathcal{C}}}
\newcommand{\Up}{U_{\mathcal{P}}}

\usepackage{ulem}

\begin{document}
\title{Generalized Berry phase for a bosonic Bogoliubov system with exceptional points}
\author{Terumichi Ohashi$^1$}
\author{Shingo Kobayashi$^{1,2}$}
\author{Yuki Kawaguchi$^1$}
\affiliation{$^1$Department of Applied Physics, Nagoya University, Nagoya 464-8603, Japan}
\affiliation{$^2$Institute for Advanced Research, Nagoya University, Nagoya 464-8601, Japan}

\date{\today}

\begin{abstract}
We discuss the topology of Bogoliubov excitation bands from a Bose-Einstein condensate in an optical lattice.
Since the Bogoliubov equation for a bosonic system is non-Hermitian, complex eigenvalues often appear and induce dynamical instability.
As a function of momentum, the onset of appearance and disappearance of complex eigenvalues is an exceptional point (EP), which is a point where the Hamiltonian is not diagonalizable and hence the Berry connection and curvature are ill-defined, preventing defining topological invariants.
In this paper, we propose a systematic procedure to remove EPs from the Brillouin zone by introducing an imaginary part of the momentum.
We then define the Berry phase for a one-dimensional bosonic Bogoliubov system.
Extending the argument for Hermitian systems, the Berry phase for an inversion-symmetric system is shown to be $\mathbb{Z}_2$.
As concrete examples, we numerically investigate two toy models 
and confirm the bulk-edge correspondence even in the presence of complex eigenvalues.
The $\mathbb{Z}_2$ invariant associated with particle-hole symmetry and the winding number for a time-reversal-symmetric system are also discussed.
\end{abstract}
\maketitle
\section{Introduction}
Topological phases of matter have attracted much attention in solid-state physics in more than one decade~\cite{RevModPhys.82.3045,RevModPhys.83.1057,Sato_2017,1803.00249}.
Although most of such studies treat Hermitian Hamiltonians, there has been growing interest in non-Hermitian systems~\cite{PhysRevB.84.153101,PhysRevB.84.205128,PhysRevLett.120.146402}.
Non-Hermitian systems exhibit unique phenomena with no counterpart in Hermitian ones, such as non-reciprocal transport~\cite{PhysRevLett.106.213901,regensburger2012parity,feng2013experimental,peng2014parity}, enhanced sensitivity~\cite{PhysRevLett.112.203901,PhysRevLett.117.110802,hodaei2017enhanced,chen2017exceptional,1805.12001,PhysRevA.95.023812}, high-performance lasers~\cite{PhysRevA.82.031801,PhysRevLett.106.093902,PhysRevLett.113.053604,Peng328,Feng972,Hodaei975}, and unconventional quantum criticality~\cite{PhysRevX.4.041001,PhysRevLett.119.190401,ashida2017parity,PhysRevLett.121.203001,1812.01213}.
Non-Hermiticity also leads to richer topological properties absent in Hermitian systems.
For example, appearance of an edge state localized at only one side of the system 
is predicted~\cite{PhysRevLett.116.133903,PhysRevLett.118.040401,PhysRevB.98.165148,PhysRevB.97.121401}.
It is also pointed out that the conventional bulk-edge correspondence breaks down in certain non-Hermitian models, and there have been great efforts to establish modified bulk-edge correspondences~\cite{PhysRevLett.118.040401,PhysRevB.97.121401,PhysRevB.99.121411,PhysRevLett.116.133903, PhysRevB.97.121401, PhysRevB.98.165148,PhysRevB.98.155430, PhysRevLett.122.076801, Xiong_2018, PhysRevLett.121.026808, PhysRevLett.118.040401,PhysRevLett.121.136802, PhysRevLett.121.086803,PhysRevB.99.121411,PhysRevB.99.081302,1810.04527,PhysRevB.99.075130,1902.10958,kawabata2019topological}.
More recently, topological classification of non-Hermitian systems has been done \cite{PhysRevX.8.031079,1812.09133,zhou2018periodic,1902.08479}.

Existence of exceptional points (EPs) is also a specific feature absent in Hermitian systems.
Due to non-Hermiticity of the system,
a non-Hermitian Hamiltonian cannot be diagonalized for a certain set of parameters that describe the Hamiltonian.
That is an EP~\cite{1751-8121-45-44-444016,0034-4885-70-6-R03,Berry2004}.
At an EP, two eigenstates coalesce and become identical to each other. 
Hence, topological invariants cannot be defined in the presence of an EP below Fermi energy, even though the system is gapfull.
So far, topological charge of an EP and deformation of a Weyl point to an exceptional ring (EPs aligned in a ring shape) accompanied with the appearance of bulk Fermi arcs has been discussed~\cite{zhen2015b,PhysRevLett.116.156803,PhysRevLett.118.045701,PhysRevB.96.045437,PhysRevB.97.041203,PhysRevB.97.075128,PhysRevB.97.220301,PhysRevLett.120.146402,Zhou1009,PhysRevA.98.042114,1810.09231,1812.02011,Zhou:19,PhysRevLett.120.146601,PhysRevB.99.041116,PhysRevB.99.041406,PhysRevB.99.041202,PhysRevB.99.081102,Zhou:19,PhysRevB.99.075130,PhysRevB.99.121101,1903.12187,1901.05047}.

In this paper, we consider topology of excitation bands from a Bose-Einstein condensate (BEC) in the presence of EPs.
We find that in some special cases EPs can be removed from the Brillouin zone (BZ) and topological invariants can be defined 
by a simple generalization of the ones defined in Hermitian systems.

In general, a non-Hermitian Hamiltonian describes an open quantum system in which loss and gain of particles coexist.
Examples include photons in nanostructures such as photonic crystals~\cite{st2017lasing,zhao2018topological,PhysRevLett.120.113901,Hararieaar4003,Bandreseaar4005}, micro cavities~\cite{RevModPhys.86.1391,jing2015optomechanically,xu2016topological},
and electric circuits~\cite{PhysRevA.84.040101,PhysRevB.97.220301}, quantum walks with non-unitary time evolution~\cite{1609.09650,xiao2017observation}, cold atomic gases with controlled loss and gain \cite{PhysRevLett.110.035302,Tomitae1701513,li2019observation,PhysRevA.99.011601,PhysRevA.87.051601,PhysRevA.90.033630}, and solid-state systems with finite quasiparticle lifetimes \cite{PhysRevLett.121.026403,PhysRevB.97.041203,1708.05841,PhysRevB.98.035141}.
Here, we note that bosonic Bogoliubov quasi-particles, which are elementary excitations from a Bose-Einstein condensate (BEC), are also described with a non-Hermitian Hamiltonian, where a BEC works as a particle bath.
The characteristic of the bosonic Bogoliubov system is that the non-Hermitian Hamiltonian always satisfies pseudo-Hermiticity~\cite{doi:10.1063/1.1418246,doi:10.1063/1.1461427,doi:10.1063/1.1489072,0305-4470-36-25-312} and particle-hole symmetry (PHS). In addition to this, we consider inversion symmetry (IS) and/or time-reversal symmetry (TRS) and discuss how topological invariants are calculated in the presence of EPs.

There are several previous works that discuss topology of bosonic Bogoliubov systems and the resulting bulk-edge correspondence, such as magnon excitations
~\cite{wang2009observation,Onose297,PhysRevLett.104.066403,PhysRevB.87.174427,PhysRevB.87.174402,PhysRevB.96.220405,PhysRevB.99.041110},
photons~\cite{PhysRevLett.93.083901,PhysRevA.78.033834,PhysRevLett.100.013904},
phonons~\cite{PhysRevLett.95.155901,PhysRevLett.96.155901},
and Bose atoms~\cite{PhysRevA.91.053621,furukawa2015excitation,PhysRevLett.115.245302,PhysRevA.93.053605,PhysRevA.88.063631,PhysRevB.93.020502,PhysRevLett.119.203204,PhysRevA.98.053617,PhysRevB.98.115135,PhysRevX.8.041031,1903.01017}.
In these works, the definition of topological invariants is extended for bosonic Bogoliubov systems, and the bulk-edge correspondence is numerically confirmed. However, despite non-Hermiticity of bosonic Bogoliubov systems, the works mentioned above discuss only the case when all eigenvalues are real. 
On the other hand, the appearance of complex eigenvalues is known to cause dynamical instability and observed in various situations~\cite{PhysRevA.59.1533,PhysRevA.62.053605,PhysRevA.64.061603,PhysRevLett.93.140406,PhysRevA.70.043610,sadler2006spontaneous}.

In this paper, we consider
the topology of bosonic Bogoliubov bands from an atomic BEC confined in an optical lattice, in particular, in the presence of complex eigenvalues. We first show that the onset of appearance and disappearance of complex conjugate eigenvalues is actually an EP.
We then introduce an imaginary part of the momentum around an EP and remove the EP from the BZ (which is defined for the real part of the momentum).
Note that this idea is distinguished from non-Bloch wave number in that we are considering not the bulk-edge correspondence~\cite{PhysRevLett.121.086803,PhysRevLett.121.136802} but bulk topology.
Due to the square root singularity of an EP, we have to be careful to choose the sign of the imaginary part of the momentum.
With this procedure, we define topological invariants in one-dimensional (1D) systems with various symmetries.
For the case of a system with IS, the band topology is characterized by a $\mathbb{Z}_2$ topological invariant, which is the Berry phase modulo $\pi$.
We can also define a $\mathbb{Z}_2$ topological invariant associated with PHS for some special cases.
When the system is invariant under time-reversal operation, we can define a winding number, which is however found to be always trivial.
As concrete examples, we consider two toy models, Kitaev-chain-like model and Su-Schrieffer-Heeger-like (SSH-like) model. 
Although these models are known to have PHS, we mainly focus on the topological property related to IS.
We numerically obtain the eigen-spectrum with open boundary condition and confirm the bulk-edge correspondence even in the presence of complex eigenvalues.
As an interesting result, we find an edge state whose energy is located in a gap of pure-imaginary bulk eigenvalues [see Fig.~\ref{fig:band_ssh}~(d)].

The organization of this paper is as follows.
In Sec.~\ref{sec:formalism}, we introduce the bosonic Bogoliubov formalism which describes the quasiparticle excitations from a condensed state. The symmetry property of the system is also given in this section.
In Sec.~\ref{sec:Mathematical framework relating to an exceptional point}, we introduce the bi-orthogonal basis and discuss the properties of EPs in the bosonic Bogoliubov system.
In particular, the extension to the complex momentum and how to remove EPs from the BZ are explained.
In Sec.~\ref{sec:topological_invariant}, we define the generalized $\mathbb{Z}_2$ topological invariant for a 1D inversion-symmetric system that can be calculated in the presence of EPs.
In Sec.~\ref{sec:other_topo_invariants}, we discuss two other topological invariants, the $\mathbb{Z}_2$ topological invariant associated with PHS and the winding number associated with chiral symmetry (CS) which is a combined symmetry of PHS and TRS.
In Sec.~\ref{sec:toy_model}, we numerically solve two toy models and confirm the bulk-edge correspondence.
In Sec.~\ref{sec:conclusion}, we conclude the paper and give some discussions. 

\section{Bogoliubov formalism \label{sec:formalism}}
Bogoliubov theory well describes quasi-particle excitations from a condensate at low temperature~\cite{Bogolyubov:1947zz,KAWAGUCHI2012253}.
In this section, we review the general properties of the Bogoliubov theory. We assume a spatially uniform or periodic system and discuss in the momentum space. However, the properties derived below, i.e., the PHS and the pseudo-Hermiticity, hold in the absence of the continuous nor discrete translational symmetries.
A general Hamiltonian is given by $\hat{\mathcal H}=\hat{\mathcal H}_0 + \hat{\mathcal V}_{\rm int}$, where
\begin{align}
	\hat{\mathcal H}_0=&\sum_{\kk,s,s'}\hat{a}^\dagger_{\kk s} (h_{\kk s s'}-\mu)\hat{a}_{\kk s'},\\
	\hat{\mathcal V}_{\rm int}=&\frac{1}{2}\sum_{\kk_1,\cdots,\kk_4}\sum_{s_1,\cdots,s_4}V_{\kk_1s_1,\kk_2s_2;\kk_3s_3,\kk_4s_4}\nonumber\\
						&\hspace{5mm}\times\hat{a}^\dagger_{\kk_1 s_1}\hat{a}^\dagger_{\kk_2 s_2}\hat{a}_{\kk_3 s_3}\hat{a}_{\kk_4 s_4}, 
	\label{eq:Vint0}
\end{align}
are the single-particle Hamiltonian and the inter-particle interaction, respectively,
with $\hat{a}_{\kk s} (\hat{a}^{\dagger}_{\kk s})$ being a bosonic operator that annihilates (creates) an atom with momentum $\kk$ in internal state $s$, and $\mu$ is the chemical potential. For the case of a periodic system, $\kk$ is a quasi-momentum and $s$ may include orbital degrees of freedom, as well as spin degrees of freedom. From the commutation relation between bosonic operators, the coefficients satisfy the following relations:
\begin{align}
	h_{\kk s s'}&=h_{\kk s' s}^*\\
	V_{\kk_1s_1,\kk_2s_2;\kk_3s_3,\kk_4s_4}&=V_{\kk_2s_2,\kk_1s_1;\kk_3s_3,\kk_4s_4} \nonumber\\
	&=V_{\kk_1s_1,\kk_2s_2;\kk_4s_4,\kk_3s_3}\nonumber\\
	&=V_{\kk_3s_3,\kk_4s_4;\kk_1s_1,\kk_2s_2}^*.
\end{align}
Due to the momentum conservation, $V_{\kk_1s_1,\kk_2s_2;\kk_3s_3,\kk_4s_4}$ is nonzero only when $\kk_1+\kk_2-\kk_3-\kk_4$ is equal to zero (a reciprocal lattice vector) in a uniform (periodic) system.

For simplicity, we consider a case when a macroscopic number of atoms are condensed in a $\kk={\bm 0}$ state. Generalization for a condensate with a nonzero momentum is straightforward. The creation and annihilation operators corresponding to the condensed state can be replaced with {\it c}-numbers, say $\hat{a}_{{\bm 0} s}\simeq \zeta_s$ and $\hat{a}_{{\bm 0} s}^\dagger\simeq \zeta_s^*$, where $\zeta_s$'s are determined so that the condensate is stationary. Expanding the Hamiltonian up to the second order of $\hat{a}_{\kk s}$ and $\hat{a}^\dagger_{\kk s}$, we rewrite the Hamiltonian in the Nambu description as follows:
\begin{align}
	\hat{\mathcal{H}}_{\rm{Bog}}
	=\frac{1}{2}\sum_{\kk}\begin{pmatrix} \tilde{\hat{\bm a}}^{\dagger}_{\kk} & \tilde{\hat{\bm a}}_{-\kk} \end{pmatrix}
	\tilde{H}(\kk)
	\begin{pmatrix}
		\hat{\bm a}_{\kk}\\
		\hat{\bm a}^{\dagger}_{-\kk}
	\end{pmatrix}
,
	\label{eq:H_Bog}
\end{align}
where $\tilde{\hat{\bm a}}_{\kk}=(\hat{a}_{\kk 1}, \hat{a}_{\kk 2},\cdots, \hat{a}_{\kk \mathcal{N}})$, $\tilde{\hat{\bm a}}^\dagger_{\kk}=(\hat{a}^\dagger_{\kk 1}, \hat{a}^\dagger_{\kk 2},\cdots, \hat{a}^\dagger_{\kk \mathcal{N}})$, $\hat{\bm a}_{\kk}=\left(\tilde{\hat{\bm a}}_{\kk}\right)^{\rm T}$, and $\hat{\bm a}_{\kk}^\dagger=\left(\tilde{\hat{\bm a}}^\dagger_{\kk}\right)^{\rm T}$ with $\mathcal{N}$ being the number of internal states. Here, $\tilde{H}(\kk)$ is a $2\mathcal{N}\times 2\mathcal{N}$ Hermitian matrix whose elements are given by
\begin{align}
	\tilde{H}(\kk)
	&=
	\begin{pmatrix}
		H^{(1)}(\kk) & H^{(2)}(\kk)\\
		[H^{(2)}(-\kk)]^* & [H^{(1)}(-\kk)]^*
	\end{pmatrix},
\end{align}
with
\begin{align}
	H^{(1)}_{ss'}(\kk)&=h_{\kk ss'}-\mu+2\sum_{s_1,s_2}V_{{\bm 0} s_1,\kk s;{\bm 0} s_2,\kk s'}\zeta^*_{s_1}\zeta_{s_2},\\
	H^{(2)}_{ss'}(\kk)&=H^{(2)}_{ss'}(-\kk)=\sum_{s_1,s_2}V_{\kk s,-\kk s';{\bm 0} s_1,{\bm 0} s_2}\zeta_{s_1}\zeta_{s_2}
\end{align}
being an $\mathcal{N}\times\mathcal{N}$ Hermitian matrix and an $\mathcal{N}\times\mathcal{N}$ symmetric matrix, respectively.

To diagonalize $\hat{\mathcal{H}}_{\rm{Bog}}$, we perform the Bogoliubov transformation:
\begin{align}
	\begin{pmatrix}
		\hat{\bm a}_{\kk}\\
		\hat{\bm a}^{\dagger}_{-\kk}
	\end{pmatrix}
	=
	\begin{pmatrix}
		u_{\kk} & v^*_{-\kk}\\
		v_{\kk} &u^*_{-\kk}
	\end{pmatrix}
	\begin{pmatrix}
		\hat{\bm b}_{\kk}\\
		\hat{\bm b}^{\dagger}_{-\kk}
	\end{pmatrix}
	=:
	T{(\kk)}
	\begin{pmatrix}
		\hat{\bm b}_{\kk}\\
		\hat{\bm b}^{\dagger}_{-\kk}
	\end{pmatrix},
\label{eq:def-Tk}
\end{align}
where $\hat{\bm b}_{\kk}=(\hat{b}_{\kk 1}, \hat{b}_{\kk 2},\cdots, \hat{b}_{\kk \mathcal{N}})^{\rm T}$ and $\hat{\bm b}^\dagger_{\kk}=(\hat{b}^\dagger_{\kk 1}, \hat{b}^\dagger_{\kk 2},\cdots, \hat{b}^\dagger_{\kk \mathcal{N}})^{\rm T}$ are the annihilation and creation operators of quasi-particles, and $T(\kk)$ is a $2\mathcal{N}\times 2\mathcal{N}$ matrix with $u_{\kk}$ and $v_{\kk}$ being $\mathcal{N}\times \mathcal{N}$ matrices.
Since both $\hat{a}_{\kk s}$ and $\hat{b}_{\kk s}$ satisfy the bosonic commutation relation, $T(\kk)$ is not a unitary matrix but a para-unitary matrix that satisfies
\begin{align}
	T^\dagger(\kk)\tau_3T(\kk)=\tau_3, 
	\label{eq:para_unitary}
\end{align}
where $\tau_{i=0,1,2,3}$ are the Pauli matrices in the Nambu space. Equation~\eqref{eq:para_unitary} is a distinctive feature of a bosonic system: the Bogoliubov transformation for a fermionic superconducting system is a unitary transformation. The Hamiltonian $\hat{\mathcal{H}}_{\rm{Bog}}$ is then diagonalized when $T(\kk)$ satisfies the following Bogoliubov equation:
\begin{align}
	H(\kk)T(\kk)
	&=
	T(\kk)
	\begin{pmatrix}
		E(\kk)&0\\
		0&-E^*(-\kk)
	\end{pmatrix},
	\label{eq:BE}
\end{align}
where
\begin{align}
	H(\kk) := \tau_3\tilde{H}(\kk)
	\label{eq:def_Hk}
\end{align}
is a non-Hermitian matrix and $E(\kk)$ is an $\mathcal{N}\times\mathcal{N}$ diagonal matrix.
In this paper, we refer to $H(k)$ defined in Eq.~\eqref{eq:def_Hk} as the bosonic Bogoliubov Hamiltonian, or simply Bogoliubov Hamiltonian, in the sense that its eigenvalues describe the excitation spectrum.
Since $H(\kk)$ is non-Hermitian, its eigenvalues are in general complex.
In Eq.~\eqref{eq:BE}, we have used PHS, i.e.,
\begin{align}
	\mathcal{C}H(\kk)\mathcal{C}^{-1}&=-H(-\kk),
	\label{eq:sym_PH}
\end{align}
where $\mathcal{C}$ is the antiunitary particle-hole operator defined by
\begin{align}
	\mathcal{C}=\tau_1K,\ \ \mathcal{C}^2=1,
	\label{eq:def_C}
\end{align}
with $K$ being the complex conjugate operator: From Eq.~\eqref{eq:sym_PH}, one can see that if $|u_n(\kk)\rangle$ is an eigenstate of $H(\kk)$ with an eigenvalue $E_n(\kk)$, $\mathcal{C}|u_n(\kk)\rangle$ is an eigenstate of $H(-\kk)$ with an eigenvalue $-E_n^*(\kk)$, obtaining Eq.~\eqref{eq:BE}. The particle-hole symmetry is a generic property of the Bogoliubov equation originating from condensation, and always holds independently from other symmetries.

We note that the Bogoliubov equation for a bosonic system is categorized to a pseudo-Hermitian eigenvalue equation. A non-Hermitian matrix $H$ is called pseudo-Hermitian when it satisfies the following relation~\cite{doi:10.1063/1.1418246,doi:10.1063/1.1461427,doi:10.1063/1.1489072,0305-4470-36-25-312}:
\begin{align}
	\eta H \eta^{-1} = H^\dagger,
\end{align}
where $\eta$ is an Hermitian matrix and referred to as a metric operator. For the case of $H(\kk)$ defined in Eq.~\eqref{eq:def_Hk}, it satisfies the pseudo-Hermiticity with  $\eta=\tau_3$, i.e.,
\begin{align}
	\tau_3 H(\kk) \tau_3^{-1} = H^\dagger(\kk).
	\label{eq:pseudo-Hermiticity}
\end{align}
From this relation, the orthonormal condition is obtained, which differs for real-eigenvalue modes and complex-eigenvalue states:
By evaluating $\langle u_n(\kk)|\tau_3 H(\kk)|u_n(\kk)\rangle=\langle u_n(\kk)|H^\dagger(\kk)\tau_3|u_n(\kk)\rangle$, we obtain $[E_n(\kk)-E_n^*(\kk)]\langle u_n(\kk)|\tau_3|u_n(\kk)\rangle=0$, which means, the normalization condition for a real-eigenvalue state can be defined as
\begin{align}
	\mbox{real } E_n(\kk):\ \ \langle u_n(\kk)|\tau_3|u_n(\kk)\rangle=1,
	\label{eq:norm_p}
\end{align}
whereas a complex-eigenvalue state is orthonormal to its conjugate
\begin{align}
	\mbox{complex } E_n(\kk):\ \ \langle u_n(\kk)|\tau_3|u_n(\kk)\rangle=0. 
	\label{eq:norm_c}
\end{align}
The above normalization condition suggests that Eq.~\eqref{eq:para_unitary} no longer holds in the presence of complex eigenvalues
and that the corresponding quasi-particles do not satisfy the bosonic commutation relation.
In such a case, however, bosonic quasi-particles can be defined by constructing linear combinations of complex eigenstates~\cite{PhysRevA.70.043610}.
We also note that when $|u_n(\kk)\rangle$ satisfies Eq.~\eqref{eq:norm_p}, the corresponding hole excitation $|u_{n'}(-\kk)\rangle=\mathcal{C}|u_n(\kk)\rangle$ has a negative norm:
\begin{align}
	\braket{ u_{n'}(-\kk)|\tau_3|u_{n'}(-\kk)}=-1.
	\label{eq:norm_n}
\end{align}
In other words, for the case of bosonic Bogoliubov equation, we can classify real-eigenvalue eigenstates into particle and hole excitations according to the sign of the ``norm", $\langle u_n(\kk)|\tau_3|u_n(\kk)\rangle$, although they describe physically the same excitation. In the following, we refer to an eigenstate satisfying Eq.~\eqref{eq:norm_p} as a positive-norm state and to that satisfying Eq.~\eqref{eq:norm_n} as a negative-norm state. 

The rest of this paper discusses topological properties of the eigen-spectrum of the Bogoliubov Hamiltonian~\eqref{eq:def_Hk}. In order to classify the system in terms of symmetries, we here summarize the related symmetry operations:
\begin{itemize}
	\item {\bf pseudo-Hermiticity:} as seen in the above, $H(\kk)$ always satisfies the pseudo-Hermiticity relation:
	\begin{align}
		\eta H(\kk)\eta^{-1}&=H^\dagger(\kk),\ \ \ \eta=\tau_3.
		\label{eq:pH}
	\end{align}

	\item {\bf particle-hole symmetry (PHS):} as a consequence of condensation, $H(\kk)$ always preserves particle-hole symmetry:
	\begin{align}
		\mathcal{C} H(\kk)\mathcal{C}^{-1}&=-H(-\kk),\ \ \ \mathcal{C}=\tau_1K.
		\label{eq:phs}
	\end{align}

	\item {\bf time-reversal symmetry (TRS):} a system preserves time-reversal symmetry when $H(\kk)$ satisfies
	\begin{align}
		\Theta H(\kk)\Theta^{-1}&=H(-\kk),
	\label{eq:TRS}
	\end{align}
	where $\Theta$ is the antiunitary time-reversal operator. Due to Bose statistics, it satisfies $\Theta^2=+1$, and therefore, there is no Kramers degeneracy.
	Since the time-reversal operation does not mix the particle and hole sectors, $\Theta$ can be written as
	\begin{align}
		\Theta = \begin{pmatrix} U & 0 \\ 0 & U^* \end{pmatrix}K
	\label{eq:Theta}
	\end{align}
	with $U$ being an $\mathcal{N}\times\mathcal{N}$ unitary matrix.

	\item {\bf chiral symmetry (CS):} a system preserves chiral symmetry when $H(\kk)$ satisfies
	\begin{align}
	\label{chiral_symmetry}
		\Gamma H(\kk)\Gamma^{-1}&=-H(\kk),
	\end{align}
	where $\Gamma$ is the chiral operator. Since the bosonic Bogoliubov Hamiltonian always preserves PHS, the presence of CS and TRS are equivalent and they are related to each other via $\Gamma=\mathcal{C}\Theta$.

	\item {\bf inversion symmetry (IS):} a system preserves the inversion symmetry when $H(\kk)$ satisfies
	\begin{align}
		\mathcal{P}H(\kk)\mathcal{P}^{-1}&=H(-\kk),
	\label{eq:IS}
	\end{align}
	where $\mathcal{P}$ is the inversion operator.
\end{itemize}

\section{Mathematical framework relating to an exceptional point}
\label{sec:Mathematical framework relating to an exceptional point}
A salient feature of a non-Hermitian system that is distinct from an Hermitian system is the existence of an EP~\cite{1751-8121-45-44-444016,0034-4885-70-6-R03,Berry2004,zhen2015b,PhysRevLett.116.156803,PhysRevLett.118.045701,PhysRevB.96.045437,PhysRevB.97.041203,PhysRevB.97.075128,PhysRevB.97.220301,PhysRevLett.120.146402,Zhou1009,PhysRevA.98.042114,1810.09231,1812.02011,Zhou:19,PhysRevLett.120.146601,PhysRevB.99.041116,PhysRevB.99.041406,PhysRevB.99.041202,PhysRevB.99.081102,Zhou:19,PhysRevB.99.075130,PhysRevB.99.121101,1903.12187,1901.05047}.
An EP is a point in a parameter space where the Hamiltonian cannot be diagonalized. In this section, we introduce some mathematical framework relating to EPs and discuss how to avoid EPs in calculation of topological invariants in bosonic Bogoliubov systems.

\subsection{Bi-orthogonal basis}
For a non-Hermitian matrix $H(\kk)$, the right and left eigenstates are defined as
\begin{subequations}
	\begin{align}
	H(\kk)\ket{\ur_{n}(\kk)}=&E_n(\kk)\ket{\ur_n(\kk)},\\
	\bra{\ul_n(\kk)}H(\kk)=&\bra{\ul_{n}(\kk)}E_n(\kk),
	\end{align}
\end{subequations}
or equivalently,
\begin{subequations}
	\begin{align}
	H(\kk)\ket{\ur_{n}(\kk)}=&E_n(\kk)\ket{\ur_n(\kk)},\\
	H^{\dagger}(\kk)\ket{\ul_{n}(\kk)}=&E^*_n(\kk)\ket{\ul_n(\kk)},
	\end{align}
\end{subequations}
where $\bra{A}$ and $\ket{A}$ are conjugate transpose of each other. In this notation, the eigenstate $|u_n(\bm k)\rangle$ appearing in Sec.~\ref{sec:formalism} should be replaced with $|\ur_n(\bm k)\rangle$. By evaluating $\langle \ul_n(\kk)|H(\kk)|\ur_m(\kk)\rangle$ and $\langle \ur_n(\kk)|H^\dagger(\kk)|\ul_m(\kk)\rangle$, one can see that two eigenstates with different eigenvalues are orthogonal to each other: $\braket{\ur_m(\kk)|\ul_n(\kk)}=\braket{\ul_m(\kk)|\ur_n(\kk)}=0	$ for $E_m\neq E_n$. In particular, when all eigenvalues are different, the following orthonormal and completeness conditions hold:
\begin{subequations}
\begin{align}
	\braket{\ur_m(\kk)|\ul_n(\kk)}=\braket{\ul_m(\kk)|\ur_n(\kk)}=\delta_{mn},\label{eq:biorth}\\
	\sum_n\ket{\ur_n(\kk)}\bra{\ul_n(\kk)}=\sum_n\ket{\ul_n(\kk)}\bra{\ur_n(\kk)}=\bm{1}.\label{eq:completeness}
\end{align}
\label{eq:biorth-completeness}
\end{subequations}

When two eigenvalues are the same for a certain $\kk$, say $E_1(\kk)=E_2(\kk)=E$,
there are two possibilities: a degenerate point or an EP. The difference is whether the non-Hermite matrix can be diagonalized or not. In general, a simplest form of a non-Hermitian matrix a Jordan normal form. For the case of a degenerate point, the $2\times 2$ block corresponding to $n=1$ and 2 eigenstates is given by
\begin{align}
	J_{\rm DP}=\begin{pmatrix} E & 0 \\ 0 & E \end{pmatrix},
\end{align}
whereas that for the case of an EP is given by
\begin{align}
	J_{\rm EP}=\begin{pmatrix} E & 1 \\ 0 & E \end{pmatrix}.
\end{align}
Here, $J_{\rm DP}$ has two linearly independent right eigenstates $(1,0)^{\rm T}$ and $(0,1)^{\rm T}$, which means, by choosing a proper basis, the whole set of the eigenstates satisfies Eq.~\eqref{eq:biorth-completeness} at a degenerate point. On the other hand, $J_{\rm DP}$ has only one right eigenstate $(1,0)^{\rm T}$ and one left eigenstate $(0,1)$. Accordingly, at an EP, two of left and right eigenstates of $H(\kk)$ are linearly dependent, $|\ul_1(\kk)\rangle=|\ul_2(\kk)\rangle$ and $|\ur_1(\kk)\rangle=|\ur_2(\kk)\rangle$, and they are orthogonal to their adjoint:
\begin{align}
	\braket{\ul_1(\kk)|\ur_1(\kk)}=\braket{\ur_1(\kk)|\ul_1(\kk)}=0,
\end{align}
which is often referred to as self-orthogonality~\cite{0034-4885-70-6-R03,1751-8121-45-44-444016}. The vanishment of the independent eigenstates means that the Bogoliubov transformation matrix $T^{\rm{R}}(k) :=(\ket{\ur_1(\kk)},\cdots,\ket{\ur_{2\mathcal{N}}(\kk)})$, which is identical to $T(\kk)$ defined in Eq.~\eqref{eq:def-Tk}, has zero determinant at an EP.

\subsection{Properties derived from pseudo-Hermiticity}
\label{sec:Properties_derived_from_pH}
For the case when the non-Hermitian Hamiltonian satisfies the pseudo-Hermiticity~\eqref{eq:pseudo-Hermiticity}, we have some more restrictions on the eigenstates. Though we use $\eta=\tau_3$ in the following discussion, the derived properties except for Eq.~\eqref{norm} hold for a generic metric operator.

First, we show that complex eigenvalues appear as a complex conjugate pair:
When $\ket{\ul_n(\kk)}$ is a right eigenstate of $H(\kk)$ with eigenvalue $E_n(\kk)$, $\tau_3\ket{\ul_n(\kk)}$ is also a right eigenstate of $H(\kk)$ with eigenvalue $E^*_n(\kk)$. It follows that when $E_n(\kk)$ is real and not degenerate, $\ket{\ur_n(\kk)}$ and $\tau_3\ket{\ul_n(\kk)}$ are identical up to a phase factor. Actually for the case of a bosonic Bogoliubov system, Eqs.~\eqref{eq:norm_p}, \eqref{eq:norm_n} and \eqref{eq:biorth} lead to
\begin{align}
\label{norm}
	\ket{\ur_n(\kk)} =\pm \tau_3\ket{\ul_n(\kk)},
\end{align}
where the plus (minus) sign in the right-hand side is for positive-norm (negative-norm) states. On the other hand, when ${\rm Im}\,E_n(\kk)\neq 0$, $\ket{\ur_n(\kk)}$ and $\tau_3\ket{\ul_n(\kk)}$ are linearly independent. In this case, there exists $m\neq n$ such that $E_m(\kk)=E_n^*(\kk)$ and $\ket{\ur_m(\kk)}\propto \tau_3\ket{\ul_n(\kk)}$. 

\begin{figure}
	\centering
	\includegraphics[clip,width=\linewidth]{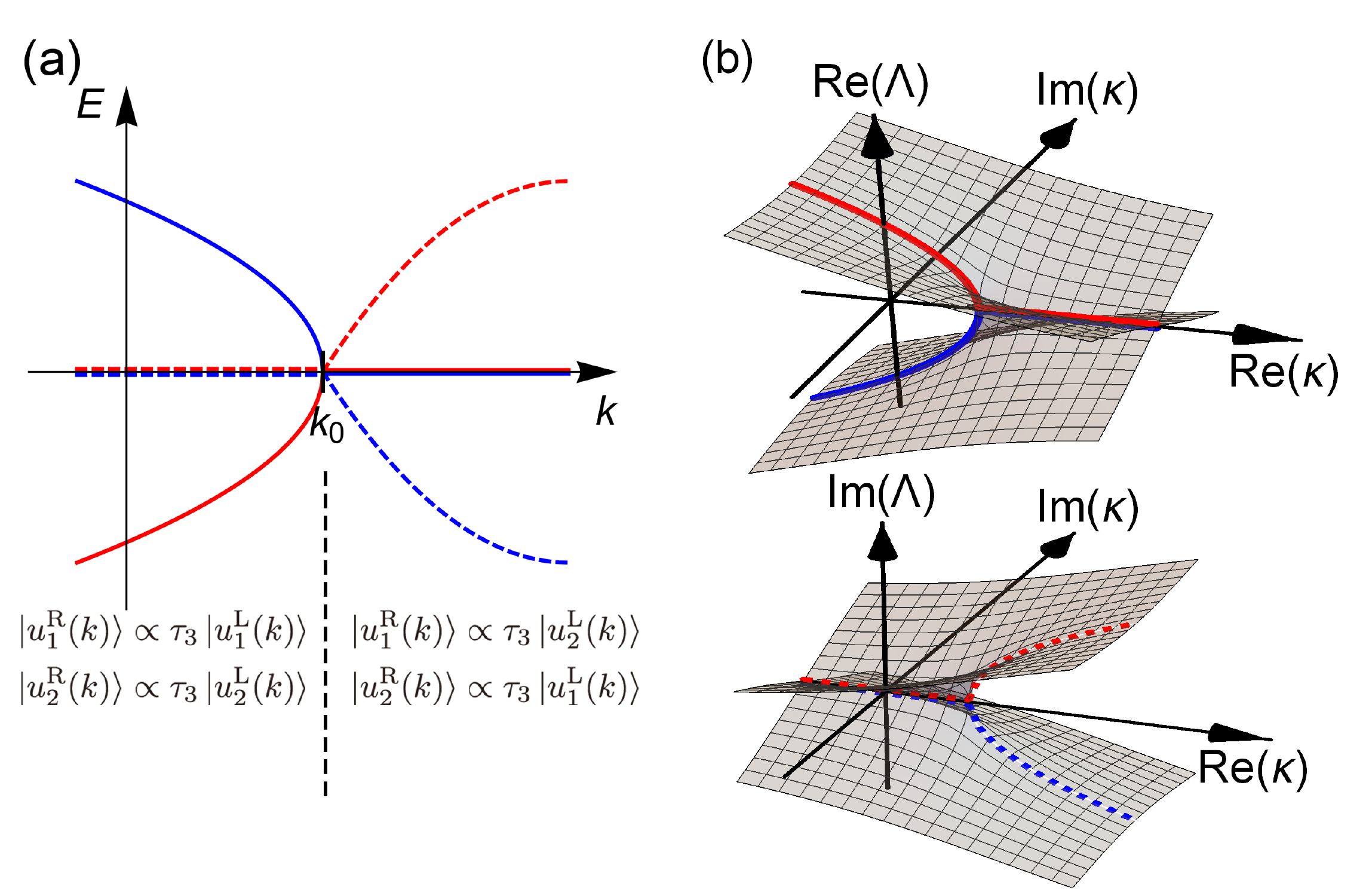}
		\caption{
		(a) Schematic energy spectrum around an EP at $k=k_0$, where two eigenvalues $E_1(k)$ and $E_2(k)$ change from two real values at $k<k_0$ to a complex conjugate pair at $k>k_0$.
		The linearly dependent partners exchange at $k=k_0$ (see text).
		(b) Two branches of a complex function $\Lambda(\kappa)=a(k_0-\kappa)^{1/2}$ which is the analytic continuations of two eigenvalues $E_{1,2}(k)$ for real $k$.
		In both figures, the red and blue solid (dashed) curves depict the real (imaginary) parts of the eigenvalues  for real $k$.
}
	\label{fig:schematic}
\end{figure}

An EP appears where two real eigenvalues coalesce and change into a pair of complex conjugate eigenvalues. To see this, we consider a one-dimensional system and assume that two eigenvalues $E_1(k)$ and $E_2(k)$ coalesce at $k=k_0$ and change from two real values at $k<k_0$ to a complex conjugate pair at $k>k_0$ [Fig.~\ref{fig:schematic} (a)]. From the above argument, linearly dependent pairs among the four eigenvectors, $\ket{\ur_1(k)}, \ket{\ur_2(k)}, \tau_3\ket{\ul_1(k)}$, and $\tau_3\ket{\ul_2(k)}$, are
\begin{subequations}
	\begin{align}
		\ket{\ur_1(k)}&\propto \tau_3\ket{\ul_1(k)},\\ 
		\ket{\ur_2(k)}&\propto \tau_3\ket{\ul_2(k)},
	\end{align}
\end{subequations}
for $k<k_0$, and
\begin{subequations}
	\begin{align}
		\ket{\ur_1(k)}&\propto \tau_3\ket{\ul_2(k)},\\
		\ket{\ur_2(k)}&\propto \tau_3\ket{\ul_1(k)},
	\end{align}
\end{subequations}
for $k>k_0$. Hence, exchange of the linearly dependent partners occurs at $k=k_0$ and $\ket{\ur_1(k)}$ and $\ket{\ur_2(k)}$ are identical at this point, i.e., $k=k_0$ is an EP. A more rigorous proof will be given in Appendix~\ref{sec:appendix_EP}. In particular, for the case of a bosonic Bogoliubov system, the two coalescing real-eigenvalue states should be a pair of positive-norm and negative-norm states, which is also explained in Appendix~\ref{sec:appendix_EP}.

We also note that since $E_1(k)-E_2(k)$ is real (pure imaginary) for $k<k_0$ ($k>k_0$), the energy spectra exhibit the square root behavior at around $k=k_0$:
\begin{align}
	E_1(k)-E_2(k)\simeq 2a \sqrt{k_0-k},
\label{eq:E1-E2}
\end{align}
where $a$ is a real number. Moreover, since EPs arise at the points where sign change of $[E_1(k)-E_2(k)]^2$ occurs, they appear as an exceptional line (surface) in a two-dimensional (three-dimensional) system.

\subsection{Extension to a complex momentum plane to avoid exceptional points}\label{subsec:avoid_ep}
Topological invariants are often defined by an integral of the Berry curvature and/or connection in the BZ,
where the system is assumed to be gapped in the domain of the integral.
In the absence of an EP, we can define the Berry curvature and connection for a non-Hermitian system, 
which will be given in Sec.~\ref{sec:topological_invariant}, and follow the argument for Hermitian systems.
The only assumption to define a topological invariant is that the energy bands are separated.
Emergence of an EP, however, causes two difficulties: (i) the fact that ${\rm det}[T^{\rm R}(k)] = 0$ makes it impossible to derive quantization of the topological charge, and 
(ii) coalescence of two energy bands makes it difficult to classify ``occupied" and ``unoccupied" bands. (Since there is no Fermi energy for a bosonic system, we define ``occupied" and ``unoccupied" bands by introducing an index $n_{\rm r}$ below which the bands are occupied.
$n_{\rm r}$ is defined such that the ``occupied" and ``unoccupied" bands do not touch, and we take the summation over ``occupied" bands to calculate the topological invariants. See, e.g., Eqs.~\eqref{IS_z2} and \eqref{eq:nu_PHS}.) Here, we show that these difficulties can be overcome when we consider a system defined on a complex $\kk$ space.
For simplicity, we assume a 1D system and use $\kappa:=k+i\epsilon\in\mathbb{C}$ with $k,\epsilon\in\mathbb{R}$ to explicitly distinguish complex ($\kappa$) and real ($k$) momentum. Generalization to higher dimensions is straightforward.

Complex momentum has also been introduced to discuss the bulk-edge correspondence in a non-Hermitian system~\cite{PhysRevLett.116.133903, PhysRevB.97.121401, PhysRevB.98.165148,PhysRevB.98.155430, PhysRevLett.122.076801, Xiong_2018, PhysRevLett.121.026808, PhysRevLett.118.040401,PhysRevLett.121.136802, PhysRevLett.121.086803,PhysRevLett.122.076801,PhysRevB.99.121411,PhysRevB.99.081302,1810.04527,PhysRevB.99.075130,1902.10958,kawabata2019topological},
where wave functions in open-boundary systems are expanded with respect to plane waves with complex wave numbers.
Our system is different from the previous works in that we consider only a bulk
and that a complex momentum is introduced just for avoiding EPs in the BZ.

We first define $H(\kappa)$ on a complex $\kappa$ plane as
\begin{align}
	H(\kappa)=
	\begin{pmatrix}
		H^{(1)}(\kappa) & H^{(2)}(\kappa) \\ -[H^{(2)}(-\kappa^*)]^* & -[H^{(1)}(-\kappa^*)]^*
	\end{pmatrix}.
	\label{eq:def_H_kappa}
\end{align}
Accordingly, the definitions for PHS, TS, CS, and IS are extended as
\begin{subequations}
\label{eq:extended_sym}
\begin{align}
	\textrm{PHS: }&\ \mathcal{C}H(\kappa)\mathcal{C}^{-1}=-H(-\kappa^*),\label{eq:extendedPHS}\\
	\textrm{TS: }&\Theta H(\kappa)\Theta^{-1}=H(-\kappa^*),\label{eq:extendedTS}\\
	\textrm{CS: }&\ \Gamma H(\kappa)\Gamma^{-1}=-H(\kappa),\label{eq:extendedCS}\\
	\textrm{IS: }&\mathcal{P}H(\kappa)\mathcal{P}^{-1}=H(-\kappa),\label{eq:extendedIS}
\end{align}
\end{subequations}
respectively.
The proof of Eq.~\eqref{eq:extended_sym} will be given in Appendix~\ref{sec:appendix_extended_sym}. The Hamiltonian $H(k_x+ik_y)$ differs from the one for a real two-dimensional system in that $\mathcal{C}$ and $\Theta$ do not change the sign of $k_y$. In this sense, $H(\kappa)$ is unphysical and introduced just for a theoretical procedure.

On the other hand, the pseudo-Hermiticity holds only on the real axis in a complex $\kappa$ plane:
\begin{align}
	\tau_3 H(\kappa) \tau_3 = H^\dagger(\kappa) \ \ \ \textrm{for}\ \ \  {\rm Im}\,\kappa=0.
	\label{eq:pseudoH_kappa}
\end{align}
This means that a complex conjugate of an eigenvalue is not necessarily an eigenvalue of $H(\kappa)$ when ${\rm Im}\,\kappa\neq 0$.
Now, we reconsider the situation discussed in Fig.~\ref{fig:schematic} (a).
Since we are interested in the square root singularity of the spectrum, we can set $[E_1(k)+E_2(k)]/2=0$ without loss of generality. Then we obtain
\begin{subequations}
\begin{align}
	E_1(k)&=a\sqrt{k_0-k},\\
	E_2(k)&=-a\sqrt{k_0-k}.
\end{align}
\end{subequations}
When $H(k)$ is analytically continued to a complex plane as $H(\kappa)$, the analytic continuations of $E_1(k)$ and $E_2(k)$ are two branches of
\begin{align}
	\Lambda(\kappa)=a(k_0-\kappa)^{1/2}.
\end{align}
From this form, one can see that the EP at $\kappa=k_0$ is isolated in the complex plane and the two eigenvalues are exchanged when one goes around the EP~\cite{1751-8121-45-44-444016}.

\begin{figure*}
\centering
\hspace{-2mm}
\includegraphics[clip,width=\linewidth]{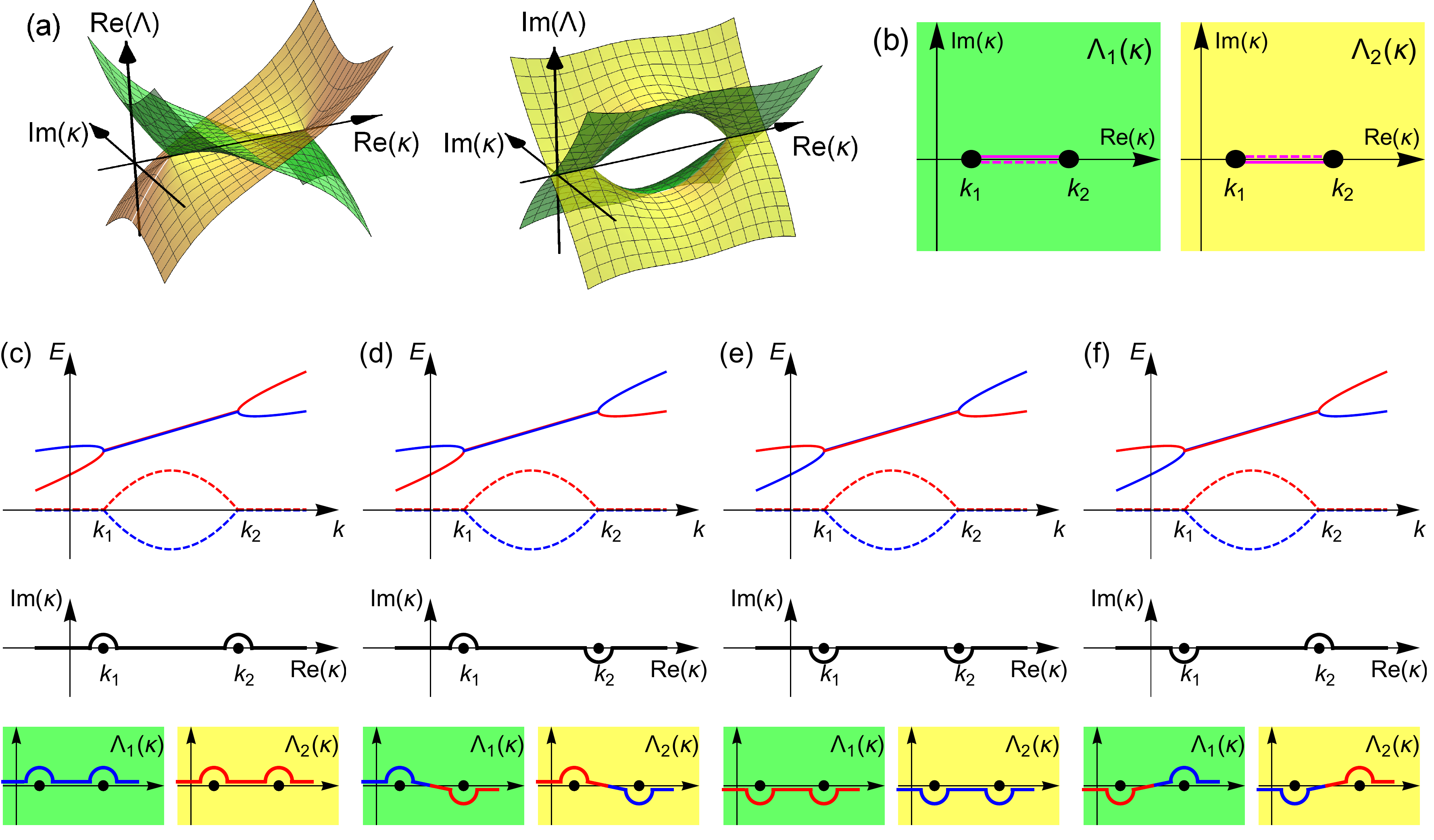}
\caption{(a) Branch structure of a square root function $\Lambda(\kappa)=\sqrt{(\kappa-k_1)(\kappa-k_2)}$, (b) the Riemann surface of $\Lambda(\kappa)$, and (c)-(f) examples of integration paths determined from how real and complex eigenvalues are continued at EPs.
In general, when two bands coalesce at $k=k_i\ (i=1,2,\cdots)$, the branch structure of the eigenvalues is the same as that of $\prod_i\sqrt{\kappa-k_i}$.
In this figure, we consider a pair of EPs at $k=k_1$ and $k_2$ ($k_1<k_2$) and assume that the eigenvalues are real (complex) for $k<k_1$ and $k>k_2$ ($k_1<k<k_2$), where $k={\rm Re}\kappa$.
The corresponding complex function $\Lambda(\kappa)$ has two branches, $\Lambda_1(\kappa)$ and $\Lambda_2(\kappa)$, whose real (imaginary) parts are depicted in the left (right) panel of (a) with green and yellow surfaces, respectively.
The Riemann surface is composed of two sheets as shown in (b), where
a branch cut is located on the real axis between the two EPs and the two branches are exchanged when we go across the cut.
In (b), the solid (dashed) magenta line in the left panel is identical to the solid (dashed) magenta line in the right panel on the Riemann surface.
The two branches satisfying $\Lambda_1(k)>\Lambda_2(k)$ for $k<k_1$ are continued to ${\rm Im}\Lambda_1(\kappa) <0 (>0)$ and ${\rm Im}\Lambda_2(\kappa)>0 (<0)$ at ${\rm Im}\kappa>0 (<0)$, and $\Lambda_1(k)<\Lambda_2(k)$ at $k>k_2$ as shown in (a).
In the top panels of (c)-(f), the real (imaginary) parts of two coalescing eigenvalues are shown with solid (dashed) lines. The red and blue colors of the lines indicate how the eigenvalues are continued. The corresponding paths in the complex $\kappa$ plane is shown in the middle panels of (c)-(f).
The bottom panels of (c)-(f) show how the red and blue bands in the top panels move in the Riemann surface.
It follows from the branch structure shown in (a) that if two eigenvalues $E_1$ and $E_2$ are continued such that the sign of $E_1(k)-E_2(k)$ for $k<k_1$ and that for $k>k_2$ are opposite, 
the integration path $\ell$ does not cross the branch cut as shown in the middle panels of (c) and (e), i.e., the red and blue bands move only in one of the two Riemann sheets.
On the other hand, if the signs of $E_1(k)-E_2(k)$ for $k<k_1$ and $k>k_2$ are the same, 
$\ell$ crosses the branch cut as shown in the bottom panels of (d) and (f).
Whether $\ell$ circumnavigates the EP at $\kappa=k_1$ in the clockwise or anti-clockwise direction is determined by the signs of ${\rm Re}(E_1-E_2)$ and ${\rm Im}E_{1,2}$.
}
	\label{fig:path}
\end{figure*}

The above argument indicates that an EP can be circumvented by taking a path in the complex plane. In the following discussions, we introduce an infinitesimal imaginary part $\epsilon(k)$, which is a function of $k$, and perform the integration along a path $\ell:\kappa(k)=k+i\epsilon(k)\ (-\pi\le  k \le \pi)$ in the complex plane:
\begin{align}\label{circumvent_integral}
	\int_{-\pi}^\pi  f(k)dk := \int_{-\pi}^\pi  f(\kappa) \frac{d\kappa}{dk} dk,
\end{align}
where $f(\kappa)$ in the right-hand side is the analytic continuation of the integrand $f(k)$ defined on the real axis except for at EPs according to Eq.~\eqref{eq:def_H_kappa}.
The path $\ell$, i.e., which path we choose in the upper- or lower-half plane at each EP, is defined such that the eigenvalues labeled in a certain manner are continuously change along the path.
The labeling rule for eigenvalues depends on what topological invariant we discuss,
and the detailed manner is given in the following sections.
Once how the real- and complex-eigenvalue states are continued at EPs is determined, we can draw a path in the complex $\kappa$ plane.
Examples are shown in Fig.~\ref{fig:path}.

\subsection{Nambu-Goldstone modes}
Before closing this section, we have to comment on EPs coming from Nambu-Goldstone modes.
Nambu-Goldstone mode is a gapless mode associated with the spontaneous breaking of a continuous symmetry due to condensation.
A reduction of ${\rm rank}\,H(\kk)$ occurs at the gap closing point, and hence, this is an EP.
However, this type of EPs can be removed by adding an infinitesimal external field that breaks the continuous symmetry.
For example, for the case of the Nambu-Goldstone phonon, which comes from a spontaneous U(1) symmetry breaking, an infinitesimal gap opens when one shifts the chemical potential in the negative direction: $\mu\to\mu-0$.
With this procedure, we treat Nambu-Goldstone modes as if it is gapped and do nothing special for this type of EPs.

\section{$\mathbb{Z}_2$ invariant associated with inversion symmetry}\label{sec:topological_invariant}
From now on, we restrict ourselves to a 1D periodic system. In particular in this section, we consider an inversion-symmetric BEC in a 1D optical lattice 
and derive a topological invariant associated with IS.

\subsection{Berry connection}
One of the fundamental field that characterizes topology of quantum states is the Berry connection. 
For the case of non-Hermitian systems, quantization of Berry phase is predicted by various authors~\cite{Berry_1985,garrison1988complex,mostafazadeh1999new,schomerus2013topologically,PhysRevB.95.174506,PhysRevLett.118.040401,weimann2017topologically,PhysRevA.97.042118,PhysRevB.97.045106,PhysRevLett.120.146402}, where the Berry connection is defined by using the bi-orthogonal basis. 
Here, we define the Berry connection matrix as
\begin{align}
	\label{berry_all}
	\mathcal{A}(k)
	&:=\frac{\mathcal{A}^{\rm{LR}}(k)+\mathcal{A}^{\rm{RL}}(k)}{2}
	,
\end{align}
where
\begin{subequations}
	\begin{align}
		\label{berry_lr}
		[\mathcal{A}^{\rm{LR}}(k)]_{mn}=i\braket{\ul_m(k)|\partial_{k}\ur_n(k)},\\
		\label{berry_rl}
		[\mathcal{A}^{\rm{RL}}(k)]_{mn}=i\braket{\ur_m(k)|\partial_{k}\ul_n(k)}.
	\end{align}
\end{subequations}
Because of the relation $[\mathcal{A}^{\rm{LR}}(k)]^{\dagger}=\mathcal{A}^{\rm{RL}}(k)$, $\mathcal{A}(k)$ is an Hermitian matrix and ${\rm tr}\mathcal{A}(k)$ is always real.
Although the four kinds of  matrices, $\mathcal{A}^{\rm{LR}}(k), \mathcal{A}^{\rm{RL}}(k), \mathcal{A}^{\rm{RR}}(k)$, and $\mathcal{A}^{\rm{LL}}(k)$, have been introduced in the context of topological phases~\cite{PhysRevLett.120.146402}, we consider only two of them since the natural inner product is defined between the right- and left-eigenstates.

It can be proved that the above defined $\mathcal{A}(k)$ satisfies the sum rule:
\begin{align}\label{sum_rule}
	\int^{\pi}_{-\pi}\frac{dk}{2\pi}\sum_n[\mathcal{A}(k)]_{nn} = N  \in \mathbb{Z},
\end{align}
without assuming any symmetry.
We note that in the presence of EPs, we define the extended Berry connection matrix just by replacing $k\in\mathbb{R}$ with $\kappa\in\mathbb{C}$
and use Eq.~\eqref{circumvent_integral}.
Equation~\eqref{sum_rule} always holds independently from how we choose the integration path.
The proof of Eq.~\eqref{sum_rule} will be given in Appendix~\ref{sec:appendix_prof_sum_rule}.

\subsection{$\mathbb{Z}_2$ topological invariant associated with inversion symmetry}
For the case when all eigenvalues are real, the diagonal terms of the  matrix defined in Eq.~\eqref{berry_all} agree with those defined in the previous works~\cite{PhysRevA.91.053621,furukawa2015excitation}:
\begin{align}
 [\mathcal{A}(k)]_{nn}=i {\rm sign}(n)\langle \ur_n(k)|\tau_3\partial_k\ur_n(k)\rangle,
\end{align}
where we have assigned positive (negative) index $n$ for positive-norm (negative-norm) states.
Following the discussion in Ref.~\cite{PhysRevA.91.053621},
it is natural to define the generalized $\mathbb{Z}_2$ topological invariant 
by using $\mathcal{A}(k)$ defined in Eq.~\eqref{berry_all} as
\begin{align}\label{IS_z2}
	(-1)^{\nu_{\rm IS}} &:= \exp\left(i\int^{\pi}_{-\pi}dk\sum_{n=1}^{n_{\rm r}}[\mathcal{A}(k)]_{nn}\right).
\end{align}
Here, $n_{\rm r}$ is a reference index defined such that the bands with $0<n\le n_{\rm r}$ and $m>n_{\rm r}$ are separated, i.e., $E_n(k)\neq E_m(k)$ for all $k$ in the BZ.
As shown below, the above defined $\nu_{\rm IS}$ takes $0$ or $1$, given by
\begin{align}\label{IS_z2_P-eigenvalue}
	(-1)^{\nu_{\rm IS}} &=\prod_{n=1}^{n_{\rm r}} \frac{\xi_n(\pi)}{\xi_n(0)},
\end{align}
where $\xi_n(k_{\rm IS}) =\pm 1$ is the eigenvalue of the inversion operator $\mathcal{P}$ for the $n$th eigenstate at the inversion-invariant momenta $k_{\rm IS}=0$ and $\pi$, i.e., $\mathcal{P}|u_n^{\rm L,R}(k_{\rm IS})\rangle=\xi_n(k_{\rm IS})|u_n^{\rm L,R}(k_{\rm IS})\rangle$.
From Eq.~\eqref{IS_z2_P-eigenvalue}, we can rewrite Eq.~\eqref{IS_z2} as
\begin{align}
	(-1)^{\nu_{\rm IS}} &:= \exp\left(i\int^{\pi}_{-\pi}dk\sum_{n=-\mathcal{N}}^{n_{\rm r}}[\mathcal{A}(k)]_{nn}\right),
\end{align}
since among the negative-index states
the numbers of even-parity and odd-parity states are the same at each $k=0$ and $\pi$, i.e., $\prod_{n=-\mathcal{N}}^{-1}\xi_n(0)=\prod_{n=-\mathcal{N}}^{-1}\xi_n(\pi)=(-1)^{\mathcal{N}/2}$.

\subsection{Labeling rule}
\label{sec:labeling_IS}
\begin{figure}
\centering
\includegraphics[width=0.9\linewidth]{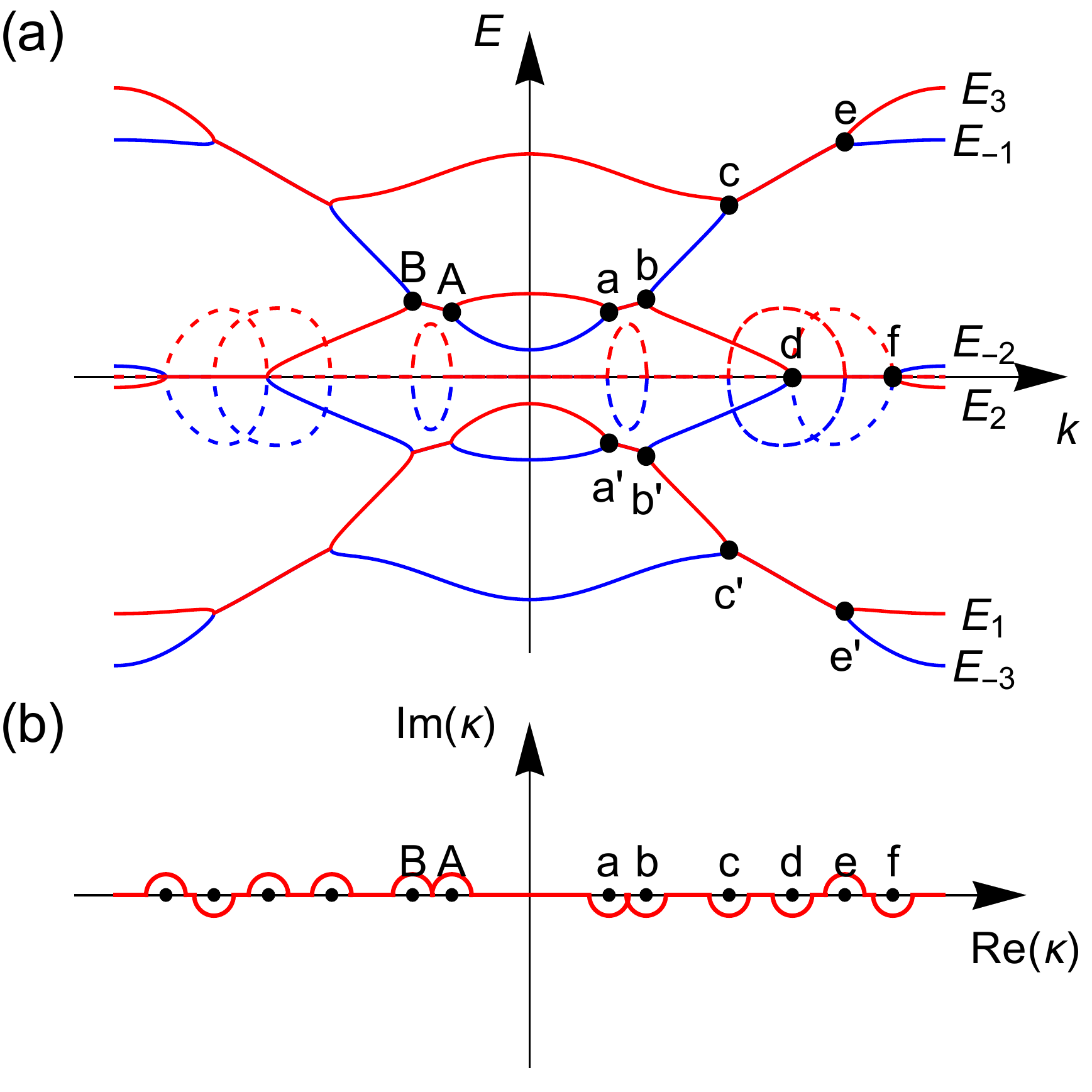}
\caption{(a) Example of a band structure and (b) the corresponding integration path. In (a), there are three positive-index (negative-index) bands labeled with $E_{1,2,3}$ ($E_{-1,-2,-3}$), which are depicted with red (blue) lines.
The real (imaginary) parts of the eigenvalues are drawn with solid (dashed) lines. 
Focusing on the EPs at {\bf a} and {\bf b}, between which a pair of complex conjugate eigenvalues appear,
the band structure is the same as Fig.~\ref{fig:path}~(e).
Due to PHS, EPs also appear at {\bf a'} and {\bf b'}.
Since we have chosen ${\rm Im}E_1>0$, the band structure around {\bf a'} and {\bf b'} is the same as {\bf a} and {\bf b}, i.e., the same as Fig.~\ref{fig:path}~(e).
On the other hand, due to the inversion symmetry, a pair of EPs appears at {\bf A} and {\bf B}, around which the band structure is the same as Fig.~\ref{fig:path}~(c).
There are another pairs of EPs at {\bf c} and {\bf e}, {\bf c'} and {\bf e'}, and {\bf d} and {\bf f}. 
The former two are in the same configuration as Fig.~\ref{fig:path}~(f), and the latter is the same as Fig.~\ref{fig:path}~(e).
The inversion symmetric counterpart of these EP pairs are respectively in the configurations of Fig.~\ref{fig:path}~(f), and Fig.~\ref{fig:path}~(c),
resulting in the integration path symmetric with respect to $\kappa=0$ as shown in (b).
Note that the integration path crosses the branch cut once (twice, which is equivalent to zero times) between the EPs at {\bf c} and {\bf e} ({\bf d} and {\bf f}).
}
	\label{fig:path_IS}
\end{figure}

To define the integration path in Eq.~\eqref{IS_z2} in the presence of EPs, we have to properly label the eigenstates.
Here, only the requirement is that the indices assigned to a pair of inversion symmetric states have the same sign, i.e., an inversion symmetric counterpart of a particle-band (hole-band) state is in a particle (hole) band.
To satisfy the requirement, we assign positive (negative) indices to real-eigenvalue positive-norm (negative-norm) states according to Eq.~\eqref{norm}. For complex-eigenvalue states, we assign a positive (negative) index to a state with ${\rm Im}E>0$ (${\rm Im}E<0$). 
Since $|\ur\rangle$ and $\mathcal{P}|\ur\rangle$, which are eigenstates of $H(k)$ and $H(-k)$, respectively, share the same eigenvalue $E$, this labeling rule satisfies our requirement. 
The positive- and negative-index states are respectively sorted in a certain manner, say in the order of ${\rm Re}E$.
Then, according to the discussion in Fig.~\ref{fig:path}, the integration path becomes symmetric with respect to $\kappa=0$, i.e., the path $\ell:\kappa(k)=k+i\epsilon(k)$ satisfies $\epsilon(-k)=-\epsilon(k)$.
For example, when the structure shown in Fig.~\ref{fig:path}~(c) [Fig.~\ref{fig:path}~(d)] appears in the $k> 0$ region, the structure of Fig.~\ref{fig:path}~(e)  [Fig.~\ref{fig:path}~(d)] should appear in the $k<0$ region due to IS.
(Note that the $k$ dependence of ${\rm Re}(E_1+E_2)$ is irrelevant in determining the integration path.)
In the case when complex eigenvalues appear at $k=0$, the band structure around $k=0$ should be those shown in Figs.~\ref{fig:path}~(d) or \ref{fig:path}~(f), for which the integration path is symmetric with respect to the origin.
An example of a more complicated band structure is given in Fig~\ref{fig:path_IS}.

\subsection{Proof of $\nu_{\rm IS}\in\mathbb{Z}_2$}
\label{sec:proof_nu_IS}
We now prove Eq.~\eqref{IS_z2_P-eigenvalue}.
The derivation is the same as the argument for topological insulators with IS~\cite{PhysRevB.76.045302,PhysRevA.91.053621}.
From IS, we obtain the following relation
\begin{align}
	&\sum_{n=1}^{n_{\rm r}}[\mathcal{A}(-\kappa)]_{nn}+\sum_{n=1}^{n_{\rm r}}[\mathcal{A}(\kappa)]_{nn}\nonumber\\
	&=-\frac{i}{2}\partial_\kappa\left[\log \,{\rm det} [\Up^{\rm{LR}}(\kappa)\Up^{\rm{RL}}(\kappa)]\right],
\label{eq:A(k)-A(-k)-2}
\end{align}
where 
$[\Up^{\rm{LR}}(\kappa)]_{mn}=\braket{u^{\rm{L}}_m(\kappa)|\mathcal{P}|u^{\rm{R}}_n(-\kappa)}$ and
$[\Up^{\rm{RL}}(\kappa)]_{mn}=\braket{u^{\rm{R}}_m(\kappa)|\mathcal{P}|u^{\rm{L}}_n(-\kappa)}$ 
are $n_{\rm r}\times n_{\rm r}$ matrices with indices $1\le m,n\le n_{\rm r}$ and satisfy $\Up^{\rm{RL}}(\kappa) \Up^{\rm{LR}}(\kappa)^{\dagger}=\bm{1}$. 
The derivation of Eq.~\eqref{eq:A(k)-A(-k)-2} will be given in Appendix~\ref{sec:Append_IS}.

Following Eq.~\eqref{circumvent_integral},
we integrate Eq.~\eqref{eq:A(k)-A(-k)-2} with respect to $k$ from $0$ to $\pi$.
Because of $\epsilon(k)$ is an odd function of $k$, the integration of the left-hand side of Eq.~\eqref{eq:A(k)-A(-k)-2} is rewritten in the simple form:
\begin{align}
    &\int_0^\pi \sum_{n=1}^{n_{\rm r}} \big[[\mathcal{A}(-\kappa(k))]_{nn}+[\mathcal{A}(\kappa(k))]_{nn}\big]\frac{d\kappa(k)}{dk} dk
\nonumber\\
    =&\int_{-\pi}^0 \sum_{n=1}^{n_{\rm r}} [\mathcal{A}(k-i\epsilon(-k))]_{nn}\left[1+i\left.\frac{d\epsilon(k')}{dk}\right|_{k'=-k}\right] dk\nonumber\\
    &+\int_0^\pi \sum_{n=1}^{n_{\rm r}} [\mathcal{A}(k+i\epsilon(k))]_{nn}\left[1+i\frac{d\epsilon(k)}{dk}\right] dk\nonumber\\
    =&\int_{-\pi}^\pi \sum_{n=1}^{n_{\rm r}} [\mathcal{A}(k+i\epsilon(k))]_{nn}\left[1+i\frac{d\epsilon(k)}{dk}\right] dk\nonumber\\
    =&\int_{-\pi}^\pi \sum_{n=1}^{n_{\rm r}} [\mathcal{A}(k)]_{nn} dk.
\end{align}
On the other hand, the integration of the right-hand side of Eq.~\eqref{eq:A(k)-A(-k)-2} is given by
\begin{align}
	&-\frac{i}{2}\int_0^\pi \partial_\kappa\left[\log \,{\rm det} [\Up^{\rm{LR}}(\kappa)\Up^{\rm{RL}}(\kappa)]\right]\frac{d\kappa}{dk}dk\nonumber\\
	=& -\frac{i}{2}\log \frac{{\rm det} [\Up^{\rm{LR}}(\pi)\Up^{\rm{RL}}(\pi)]}{{\rm det} [\Up^{\rm{LR}}(0)\Up^{\rm{RL}}(0)]}.
\label{eq:is_Aint}
\end{align}
Note that since $H(k_{\rm IS})$ with $k_{\rm IS}=0$ and $\pi$ commutes with the inversion operator $\mathcal{P}$, $\Up^{\rm{LR}}(k_{\rm IS})$ and $\Up^{\rm{RL}}(k_{\rm IS})$ are diagonal matrices and identical to each other. The diagonal elements $\xi_n(k_{\rm IS})$ are the eigenvalues of $\mathcal{P}$: $\mathcal{P}|u_n^{\rm R}(k_{\rm IS})\rangle = \xi_n(k_{\rm IS})|u_n^{\rm R}(k_{\rm IS})\rangle$ with $\xi_n(k_{\rm IS})=1$ or $-1$. As a result, we obtain
\begin{align}
	\int_{-\pi}^\pi dk \sum_{n=1}^{n_{\rm r}}[\mathcal{A}(k)]_{nn} = -i\log \prod_{n=1}^{n_{\rm r}}\frac{\xi_n(\pi)}{\xi_n(0)},
\end{align}
or equivalently,
\begin{align}
	\exp\left(i\int_{-\pi}^\pi dk \sum_{n=1}^{n_{\rm r}}[\mathcal{A}(k)]_{nn}\right) = \prod_{n=1}^{n_{\rm r}}\frac{\xi_n(\pi)}{\xi_n(0)}.
\label{eq:is_exp}
\end{align}
Since the right-hand side of Eq.~\eqref{eq:is_exp} takes 1 or $-1$, $\nu_{\rm IS}$ defined in Eq.~\eqref{IS_z2_P-eigenvalue} takes 0 or 1.

\section{Other topological invariants}
\label{sec:other_topo_invariants}
In this section, we consider topological invariants associated with PHS and CS.
Unfortunately, the definition for the former is subtle, whereas the latter is shown to be always trivial.

\subsection{$\mathbb{Z}_2$ topological invariant associated with particle-hole symmetry}
\label{subsec:topological_invariant_phs}
Suppose that all eigenstates, including complex-eigenvalue states, are classified into particle and hole excitations and assigned positive and negative indices, respectively.
Then, we can divide ${\rm tr}\mathcal{A}(k)$ into the contributions from the positive-index and negative-index bands as
\begin{subequations}
\label{berry_phs}
	\begin{gather}
		\label{PHS_sum2}
		{\rm tr}\mathcal{A}(k) = A_{+}(k) + A_{-}(k),\\
		\label{a_p}
		A_{+}(k) := \sum^{\mathcal{N}}_{n=1}[\mathcal{A}(k)]_{nn},\\
		A_{-}(k) := \sum^{-1}_{n=-\mathcal{N}}[\mathcal{A}(k)]_{nn}.
	\end{gather}
\end{subequations}
By regarding the negative-index (positive-index) states as occupied (unoccupied) states,
the topological invariant is defined as the Berry phase of the negative index bands:
\begin{align}
    (-1)^{\nu_{\rm PHS}}:=\exp\left(i\int_{-\pi}^\pi dk A_-(k)\right).
\label{eq:nu_PHS}
\end{align}
Below, we give the labeling rules for eigenstates
and prove $\nu_{\rm PHS}=0$ or $1$ for some special cases.

\subsubsection{Labeling rule}
Since we have to classify particle and hole bands, the requirement is that a pair of particle-hole conjugate states has indices with opposite signs.
The classification for real-eigenvalue states based on Eq.~\eqref{norm} satisfies this requirement.
When complex-eigenvalue states exist,
four complex-eigenvalue states with eigenvalues $E, E^*, -E$ and $-E^*$ appear all together, which are related to each other via $\mathcal{C}$ and $\tau_3$ as summarized in Table~\ref{tab:PHS_labeling}.
As shown in Table~\ref{tab:PHS_labeling}, when $E$ is an eigenvalue with a positive index, $E^*$ and $-E^*$ ($-E$) should be assigned negative indices (a positive index).
To be consistent with the above observation, we classify complex-eigenvalue states
such that a state with $k\, {\rm Im}E>0$ ($<0$) has a positive (negative) index.
It follows that in the case when complex eigenvalues appear at $k=0$ or $\pi$,
we fail to assign indices, which means that we cannot define $\nu_{\rm PHS}$.
\begin{table}
\begin{tabular}{rcrc}
 state~~ &  Hamiltonian & eigenvalue~~~ & assigned index \\ \hline
 $|\ur\rangle$ &  $H(k)$ &  $E\ =\ \ \,a+ib$ & $+$\\
 $\tau_3|\ul\rangle$ & $H(k)$ & $E^*=\ \ \,a-ib$ & $-$\\
 $\mathcal{C}|\ur\rangle$ & $H(-k)$ & $-E^*=-a+ib$ & $-$\\
 $\mathcal{C}\tau_3|\ul\rangle$ & $H(-k)$ & $-E\ =-a-ib$ & $+$\\ \hline
\end{tabular}
\caption{Relationship between a positive-index state $|\ur\rangle$, which is a right eigenstate of $H(k)$ with an eigenvalue $E=a+ib\ (a,b\in\mathbb{R})$, and other states related via symmetry operations.
Since a pair of positive-norm and negative-norm real-eigenvalue states changes into a pair of complex conjugate states (see Sec.~\ref{sec:Properties_derived_from_pH} and Appendix~\ref{sec:appendix_EP}),  $\tau_3|\ul\rangle$ belonging to an eigenvalue $E^*=a-ib$ is assigned a negative index.
The particle-hole conjugate states $\mathcal{C}|\ur\rangle$ and $\mathcal{C}\tau_3|\ul\rangle$ are the right eigenstate of $H(-k)$ and assigned indices with opposite signs to $|\ur\rangle$ and $\tau_3|\ul\rangle$, respectively.
}
\label{tab:PHS_labeling}
\end{table}

To summarize the above argument, we label eigenstates as follows:
\begin{enumerate}
\item Check whether complex eigenvalues exist at $k=0$ or $\pi$. If they exist, we cannot define $\nu_{\rm PHS}$.
\item At each $k$ in the BZ except for EPs, classify all eigenvalues into positive- and negative-index states:
\begin{enumerate}
\item Real-eigenvalue states are classified according to Eq.~\eqref{norm}.
\item Complex-eigenvalue states are classified such that a state with $k\, {\rm Im}E>0$ ($<0$) has a positive (negative) index.
\end{enumerate}
\item Sort each set of positive- and negative-index states, for example,
in ascending order of ${\rm Re}E$ and ${\rm Im}E$.
\end{enumerate}
A path for the integration in Eq.~\eqref{eq:nu_PHS} in the complex $\kappa$ plane is determined to be consistent with the above defined band structure.
As we will see below, however, $\nu_{\rm PHS}$ is shown to belong to $\mathbb{Z}_2$ only when the integration path can be continuously deformed to a path with a fixed imaginary part, i.e., $\ell_0:\kappa(k)=k+i\epsilon_0$ with a real constant $\epsilon_0$.
Otherwise, we cannot prove $\nu_{\rm PHS}\in \mathbb{Z}_2$. (It might be $\mathbb{Z}_2$, but we have not proved.)
It follows that the band structure around all EPs should be in the configuration of Fig.~\eqref{fig:path}~(c) or Fig.~\eqref{fig:path}~(e)
for $\nu_{\rm PHS}$ to proved to be in $\mathbb{Z}_2$.

\subsubsection{Proof of $\nu_{\rm PHS}\in\mathbb{Z}_2$}
Using the extended PHS~\eqref{eq:extendedPHS}, we can prove the following relation:
\begin{align}
	A_+(-\kappa^*) - A_-(\kappa) =-\partial_\kappa\theta_\mathcal{C}(\kappa),
\label{eq:A_+-A_-}
\end{align}
where $\theta_\mathcal{C}(\kappa):={\rm arg}[ {\rm det}\Uc^{\rm RL}(\kappa)]$ with $[\Uc^{\rm{RL}}(\kappa)]_{mn}:=\braket{u^{\rm{R}}_m(\kappa)|\mathcal{C}|u^{\rm{L}}_n(-\kappa^*)}$
 being an $\mathcal{N}\times\mathcal{N}$ matrices defined for $m<0$ and $n>0$.
The detailed derivation is given in Appendix \ref{prof_phs}.
We integrate Eq.~\eqref{eq:A_+-A_-} along the path $\ell: \kappa(k)=k+i\epsilon(k)\ (-\pi\le k\le \pi)$, obtaining 
\begin{align}
		&\int_{-\pi}^{\pi}\frac{dk}{2\pi}\frac{d\kappa(k)}{dk}\left[A_+\left(-k+i\epsilon(-k)\right)-A_-\left(k+i\epsilon(k)\right)\right]\nonumber\\
		&= -\int_\ell \frac{d\kappa}{2\pi} \partial_{\kappa}\theta_\mathcal{C}(\kappa)= M\in \mathbb{Z},
\label{eq:A_+-A_-_complexplane}
\end{align}
where $M$ is the number of times $-\det U^{\rm RL}_\mathcal{C}(\kappa)$ goes around the origin of the complex plane as $\kappa$ changes along the path $\ell$.
We note that if $\epsilon(k)>0$ or $\epsilon(k)<0$ for all $k$ in the Brillouin zone and we can continuously deform the path $\ell$ to $\ell_0: \kappa(k)=k+i\epsilon_0$, Eq.~\eqref{eq:A_+-A_-_complexplane} reduces to
\begin{align}\label{PHS_berry_int}
		\int_{-\pi}^{\pi}\frac{dk}{2\pi}[A_+(k+i\epsilon_0)-A_-(k+i\epsilon_0)]
		= M\in \mathbb{Z}.
\end{align}
On the other hand, Eq.~\eqref{sum_rule} is rewritten in terms of $A_{\pm}$ as
\begin{align}
		\int_{-\pi}^{\pi}\frac{dk}{2\pi}[A_+(k)+A_-(k)]
		= N\in \mathbb{Z}.
\end{align}
Hence, from the above equations,
one can see that the Berry phase of the negative-index bands is quantized in units of $1/2$:
\begin{align}
	\int_{-\pi}^{\pi}\frac{dk}{2\pi}A_-(k)
	= \frac{N-M}{2}=\frac{N'}{2}. 
\end{align}
In general, Berry phase changes by an integer under a unitary gauge transformation. In the present case, the Berry phases of the positive-index and negative-index bands, $\int A_\pm(k)dk/(2\pi)$, independently change by integers since the positive-index and negative-index bands are not mixed under the gauge transformation due to the above introduced labeling rule.
As a result, the topological invariant is defined by $\nu_{\rm PHS}=N'$ mod $2=0$ or $1$, or equivalently,
$\nu_{\rm PHS}$ defined in Eq.~\eqref{eq:nu_PHS} belongs to $\mathbb{Z}_2$.

\subsection{Winding number associated with chiral symmetry}
\label{sec:winding_number_CS}
It is shown in Ref.~\cite{PhysRevB.84.205128} that a non-Hermitian Hamiltonian in 1D is classified by a winding number when the Hamiltonian satisfies pseudo-anti-Hermiticity $\eta H(k){\eta}^{-1}=-H(k)^{\dagger}$.
Following the argument in Ref.~\cite{PhysRevB.84.205128}, we show below that a winding number can be defined when the system preserves TRS, which means the system also preserves CS. However we also show that the winding number is always trivial in the present system. In other words, there is no topological phase protected by CS in 1D bosonic Bogoliubov system.
Since we use Hermitian part of the non-Hermitian Hamiltonian (see below), 
existence of EPs is not a matter.

Assuming a time-reversal symmetric system, we start from pseudo-Hermiticity~\eqref{eq:pseudo-Hermiticity} and CS ~\eqref{chiral_symmetry},
and obtain pseudo-anti-Hermiticity:
\begin{align}
	(\tau_3\Gamma)H(k)(\tau_3\Gamma)^{-1}=-H(k)^{\dagger}.
\end{align}
From such $H(k)$, we can construct an Hermitian Hamiltonian
\begin{align}
	\label{Hamiltonian_0}
	H_{0}(k):=\frac{H(k)+H(k)^{\dagger}}{2},
\end{align}
which satisfies the CS
\begin{subequations}
\begin{align}
\label{CS_non_Hermitian}
	&\tilde{\Gamma} H_0(k){\tilde{\Gamma}}^{-1}=-H_0(k),\\
	&\tilde{\Gamma}:=i\tau_3\Gamma,\ \ \tilde{\Gamma}^2=+1.
\label{eq:def_tGamma}
\end{align}
\end{subequations}
Since $H_0(k)$ is an Hermitian Hamiltonian, we can use the definition of the winding number for an Hermitian Hamiltonian~\cite{PhysRevB.84.205128}, i.e., 
\begin{align}\label{def:winding_number}
	w
	:=&\frac{1}{4\pi i} \int^{\pi}_{-\pi}dk\mbox{tr}[\tilde{\Gamma} {H_0(k)^{-1}}\partial_{k}H_0(k)]\\
	=&\frac{1}{2\pi}\mbox{Im}\Big[\int^{\pi}_{-\pi}dk\partial_{k}\mbox{ln}[\mbox{det} \,\tilde{q}(k)]\Big] \in \mathbb{Z},
\end{align}
where $\tilde{q}(k)$ is defined by 
\begin{align}
	U_{\tilde{\Gamma}}^\dagger H_0(k)U_{\tilde{\Gamma}}
	=
	\begin{pmatrix}
		0 & \tilde{q}(k)\\
		\tilde{q}(k)^{\dagger} & 0
	\end{pmatrix}
\end{align}
with $U_{\tilde{\Gamma}}$ being the unitary matrix that diagonalizes $\tilde{\Gamma}$.
The winding number corresponds to the number of times ${\rm det}\,\tilde{q}(k)$ goes around the origin in the complex plane as one goes through the 1D BZ.

We note that although the winding number can be defined for a 1D bosonic Bogoliubov Hamiltonian, it is shown to be always trivial.
This is because of the property of the metric operator $\tau_3$ that it commutes with $H_0(k)$ and anti-commute with $\tGamma$:
\begin{align}
[\tau_3,H_0(k)]&=0,\\
\{\tau_3,\tGamma\}&=0,
\end{align}
where we have used Eqs.~\eqref{eq:def_C}, and \eqref{eq:Theta} for the second equation.
From the above commutation relations, we can rewrite the winding number as
\begin{align*}
w	&= \frac{1}{4\pi i} \int^{\pi}_{-\pi}dk\mbox{tr}[\tGamma H_0(k)^{-1} \partial_{k}H_0(k)]\\
	&= \frac{1}{4\pi i} \int^{\pi}_{-\pi}dk\mbox{tr}[\tau_3 \tGamma \tau_3^{-1} \tau_3 H_0(k)^{-1} \tau_3^{-1}\partial_{k}\tau_3H_0(k)\tau_3^{-1}]\\
	&=-\frac{1}{4\pi i} \int^{\pi}_{-\pi}dk\mbox{tr}[\tGamma H_0(k)^{-1} \partial_{k} H_0(k)]\\
	&=-w,
\end{align*}
which means $w=0$.

\section{Toy models of non-trivial topological bands\label{sec:toy_model}}
In this section, we calculate topological invariants defined in the previous sections and numerically investigate the bulk-edge correspondence for two toy models: the one is a Kitaev-chain-like model and the other is a SSH-like model. The former preserves IS and breaks TRS, whereas the latter preserves the both symmetries.

\subsection{Kitaev-chain-like model}
\label{sec:toy_model_kitaev}
We consider a pseudo-spin 1/2 Bose gas in a 1D optical lattice with spin-orbit interaction (SOI) under a magnetic field.
SOI in cold atomic systems can be implemented by using laser induced synthetic gauge fields~\cite{lin2011spin,PhysRevLett.107.255301}.
The Hamiltonian $\mathcal{\hat{H}}$ of this system is given by
\begin{align}
	\label{Hamiltonian1}
	\mathcal{\hat{H}}
	=\sum_{j}\Big[
	&-t(\hat{a}^{\dagger}_{j,\uparrow}\hat{a}_{j+1,\uparrow}-\hat{a}^{\dagger}_{j,\downarrow}\hat{a}_{j+1,\downarrow} + \mbox{H.c.})\nonumber\\
	&+h(\hat{a}^{\dagger}_{j,\downarrow}\hat{a}_{j,\downarrow}-\hat{a}^{\dagger}_{j,\uparrow}\hat{a}_{j,\uparrow}) \nonumber\\
	&+\lambda(\hat{a}^{\dagger}_{j,\uparrow}\hat{a}_{j+1,\downarrow}-\hat{a}^{\dagger}_{j,\uparrow}\hat{a}_{j-1,\downarrow}+\mbox{H.c.})\nonumber\\
	&+\frac{g}{2}(\hat{a}^{\dagger}_{j,\uparrow}\hat{a}^{\dagger}_{j,\uparrow}\hat{a}_{j,\uparrow}\hat{a}_{j,\uparrow}
	 + \hat{a}^{\dagger}_{j,\downarrow}\hat{a}^{\dagger}_{j,\downarrow}\hat{a}_{j,\downarrow}\hat{a}_{j,\downarrow} )\Big].
\end{align}
where $\hat{a}_{j,\sigma}^\dagger$ ($\hat{a}_{j,\sigma}$) is the creation (annihilation) operator of an atom with spin $\sigma=\uparrow, \downarrow$ at the $j$th site, and
$t,h,\lambda$, and $g$ are all real and denote the hopping amplitude, an external magnetic field, strength of the SOI, and the intra-species interaction (which is assumed to be positive and the same for up and down spins), respectively. We neglect the inter-species interaction for the sake of simplicity.
\begin{figure}
	\centering
		\includegraphics[clip,width=7.0cm]{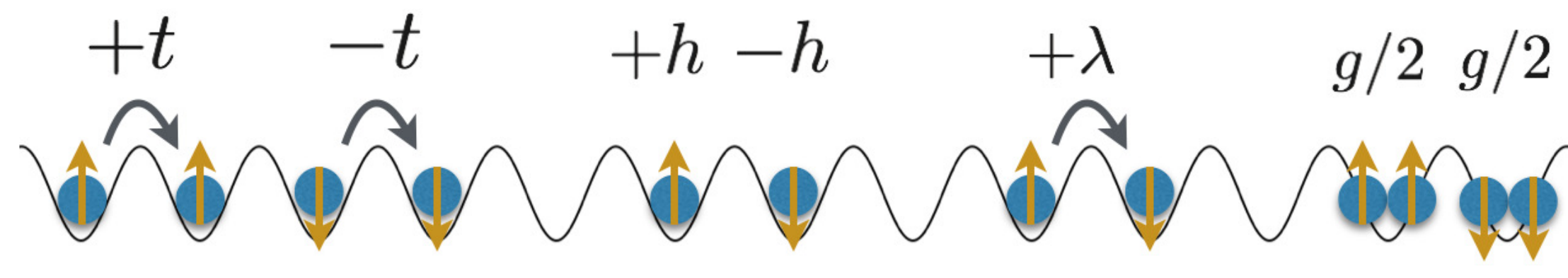}
		\caption{
		Schematic of the Kitaev-chain-like model given by Eq.~\eqref{Hamiltonian1}.
		Pseudo-spin-1/2 Bose atoms are loaded into a 1D optical lattice.
		Here, $t$ ($-t$) is the hopping amplitude for spin up (down) atoms, $h$ is the external magnetic field, $\lambda$ is the strength of the SOI, and $g$ is the intra-species interaction.
		}
	\label{kitaev_model}
\end{figure}

In fermionic systems, the Kitaev chain, that is a 1D chain of fermions with a spin-polarized superconducting {\it p}-wave paring, is the simplest model of a topological superconductor~\cite{1063-7869-44-10S-S29}.
We call the Hamiltonian~\eqref{Hamiltonian1} as the Kitaev-chain-like model because 
the system is essentially the same as the Rashba nanowire with a proximity induced {\it s}-wave paring~\cite{PhysRevLett.105.077001,PhysRevLett.105.177002}, which is known to be equivalent to the Kitaev chain,
in the sense that there exist the SOI, an external field, and the superconducting paring which corresponds to the condensate in the present case.
The previous work~\cite{PhysRevA.91.053621}  has defined the $\mathbb{Z}_2$ topological invariant associated with IS for the model of Eq.~\eqref{Hamiltonian1} and numerically confirmed the bulk-edge correspondence. Here, we use the generalized definition for the topological invariants and examine the bulk-edge correspondence in the presence of complex eigenvalues.

\begin{figure}
    \centering
	    	\includegraphics[width=8.6cm]{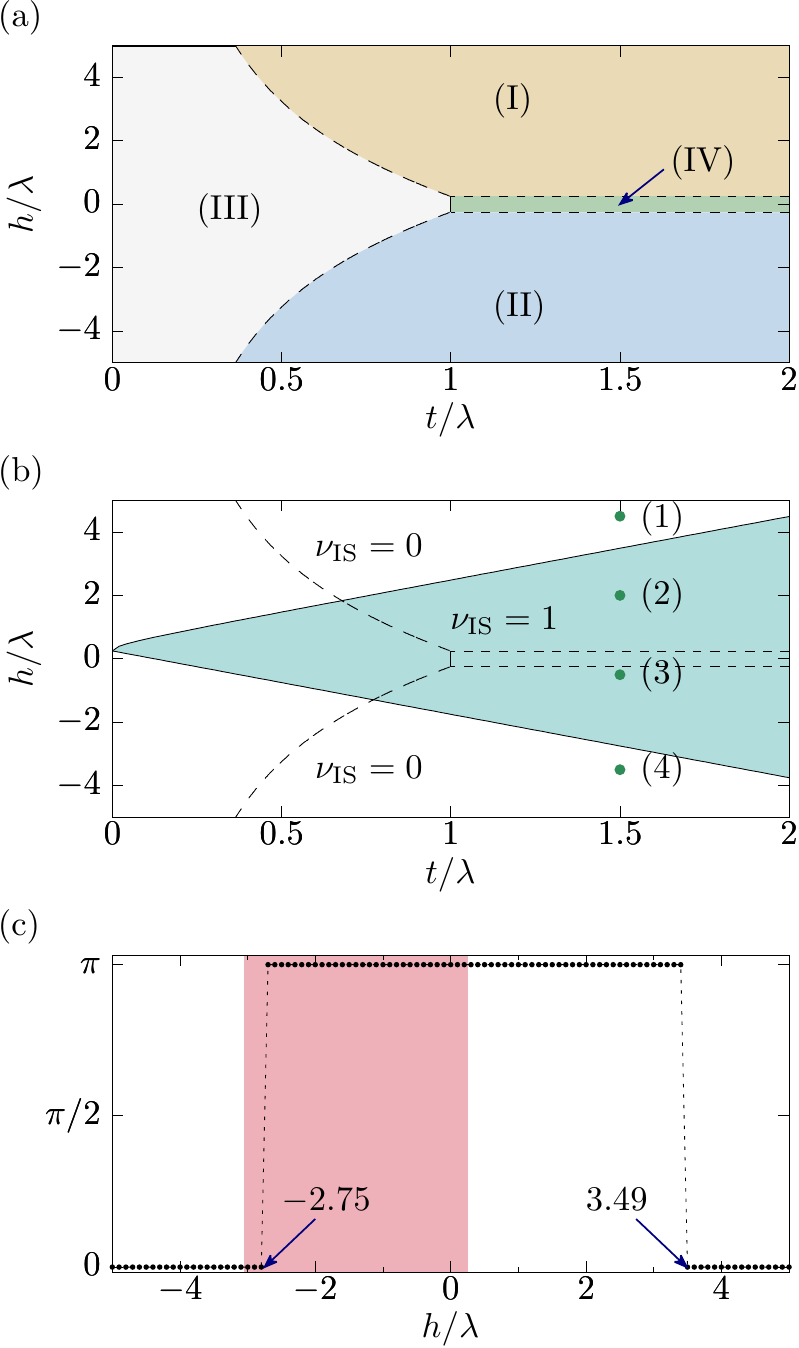}
   	\caption{
   	(a) Grand-state phase diagram of the Kitaev-chain-like model given by Eq.~\eqref{Hamiltonian1} numerically calculated for $g\rho/\lambda=0.5$.
   	There are four phases labeled with (I)-(IV). Phase I (II) is the condensation of up-spin (down-spin) atoms, Phase III is the stripe phase, 
   	and Phase IV is a superposition state of the order parameters in Phases I and II (see text).
   	We prepare a BEC in Phase I and calculate the Bogoliubov bands with changing $t/\lambda$ and $h/\lambda$.
   	When these parameters are outside of the region (I) but not so far from the phase boundary, complex eigenvalues appear in the excitation spectrum.
   	(b) Topological phase diagram for the Bogoliubov bands from a BEC in Phase I. The dashed lines depict the phase boundary shown in (a).
    Topological phase transition points are given by $h=g\rho/2+\sqrt{2t(2t+g\rho)}$ and $h=g\rho/2-2t$,
   	and the $\mathbb{Z}_2$ topological invariant $\nu_{\rm IS}$ becomes non-trivial in the cyan-colored region.
	The points labeled with (1)-(4) indicate the parameters used in the calculation of Figs. \ref{fig:band_kitaev}~(a)-\ref{fig:band_kitaev}~(d), respectively.
	(c) Berry phase as a function of $h/\lambda$ at $g\rho/\lambda=0.5$ and $t/\lambda=1.5$ numerically calculated by means of the Wilson loop method.
	The value discontinuously jumps between $0$ and $\pi$ at $h/\lambda=-2.75$ and $3.49$, which coincide with the phase transition points shown in (b).
    Complex eigenvalues appear in the red-colored region, indicating that the Berry phase and hence $\nu_{\rm IS}$ are well-defined even in the presence of complex eigenvalues.
    }
    \label{fig:gs_topo_kitaev}
\end{figure}

Assuming a superfluid phase, we obtain the mean-field ground-state phase diagram of the Kitaev-chain-like model Eq.~\eqref{Hamiltonian1} as shown in Fig~\ref{fig:gs_topo_kitaev} (a), which is calculated for $g\rho/\lambda=0.5$ with $\rho$ being the mean number density of atoms.
There are four phases: Phase I (II) is the condensation of spin-up (spin-down) atoms into the $k=0$ ($k=\pi$) state. Phase III is the so-called the stripe phase~\cite{PhysRevLett.105.160403,lin2011spin}, where atoms are condensed into a superposition state of momentum $k(\neq 0)$ and $-k$ states with opposite spin direction. 
Phase IV is a superposition of the order parameters in Phases I and II. Phase IV arises because we have neglected the inter-species interaction.
The detailed description for Phases I, II, and III is given in Ref.~\cite{PhysRevA.91.053621}.

Below, we prepare a BEC in Phase I, for which the order parameter is given by $(\langle \hat{a}_{j,\uparrow}\rangle,\langle \hat{a}_{j,\downarrow}\rangle)=(\sqrt{\rho},0)$. We then calculate the Bogoliubov excitation spectrum with changing $h/\lambda$ and $t/\lambda$.
Performing a Fourier transform $\hat{a}_{j,\sigma}=(1/\sqrt{L})\sum_k \hat{a}_{k,\sigma}e^{ikj}$ with taking the lattice constant to be unity and $L$ being the number of lattice sites, the Bogolibov equation in the momentum space is given by
\begin{align}
	H(k)T(k)&=T(k)
	\begin{pmatrix}
		E(k)&0\\
		0&-E^*(k)
	\end{pmatrix}
	,\label{eq:Hbog_kitaev}\\
	H(k)&=
	\begin{pmatrix}
		[H^{(1)}(k)]^* & H^{(2)}\\
		-H^{(2)*} & -[H^{(1)}(-k)]^*
	\end{pmatrix}
	,
	\\
	H^{(1)}(k)&=
	\begin{pmatrix}
		-2t\cos k -h +2g\rho -\mu & 2i\lambda\sin k\\
		-2i\lambda\sin k & 2t\cos k +h -\mu
	\end{pmatrix}
	,
	\\
	H^{(2)}&=
	\begin{pmatrix}
		g\rho & 0\\
		0 &0
	\end{pmatrix}
	,
\end{align}
where $\mu=-h-2t+g\rho$ is the chemical potential for the prepared state.
When the values of $h/\lambda$ and $t/\lambda$ are not in the region (I) of Fig.~\ref{fig:gs_topo_kitaev}~(a) but not so far from the phase boundary, the initially prepared state becomes dynamically unstable and complex eigenstates appear, which is the case we are interested in.

The above defined $H(k)$ satisfies pseudo-Hermiticity, PHS, and IS:
\begin{align}
    \begin{array}{rrlrl}
	\textrm{pseudo-H:}& \eta H(k){\eta}^{-1}&=\ \ {H(k)}^{\dagger},&\eta&=\tau_3,\\[1.5mm]
	\textrm{PHS:}& \mathcal{C} H(k)\mathcal{C}^{-1}&=-H(-k),\ &\mathcal{C}&=\tau_1K,\\[1.5mm]
	\textrm{IS:}& \mathcal{P} H(k)\mathcal{P}^{-1}&=\ \ H(-k),\ &\mathcal{P}&=\sigma_3,
    \end{array}
\end{align}
where $\sigma_{i=0,1,2,3}$ ($\tau_{i=0,1,2,3}$) is the Pauli matrices in the spin (Nambu) space.
According to the discussions in Sec.~\ref{sec:topological_invariant}, we can define the $\mathbb{Z}_2$ topological invariant $\nu_{\rm IS}$ associated with IS. Reference~\cite{PhysRevA.91.053621} showed for the case when all eigenvalues are real that $\nu_{\rm IS}$ takes $0$ and $1$ depending on the parameters and that an edge state arises when $\nu_{\rm IS}=1$. We have numerically confirmed that this result holds even when complex eigenvalues exist.

We calculate $\nu_{\rm IS}$ with $n_{\rm r}=1$ using Eq.~\eqref{IS_z2_P-eigenvalue}. At $k=0$ and $\pi$, the Bogoliubov equation is divided into spin-up $(u_\uparrow,v_\uparrow)$ and spin-down $(u_\downarrow,v_\downarrow)$ sectors, for which the eivenvalues of $\mathcal{P}=\sigma_3$ are given by $1$ and $-1$, respectively. The corresponding positive-index eigenvalues, which are all real, are given by
\begin{align}
    &E_\uparrow(0)=0,\\
    &E_\downarrow(0)=4t+2h-g\rho, 
\end{align}
for $k=0$ and
\begin{align}
    &E_\uparrow(\pi)=2\sqrt{2t(2t+g\rho)},\\
    &E_\downarrow(\pi)=2h-g\rho,
\end{align}
for $k=\pi$. Labeling the above eigenvalues such that $E_1<E_2$, we can divide the parameter space of Fig.~\ref{fig:gs_topo_kitaev} (a) into the following three regions:
\begin{subequations}
\begin{align}
{\rm (i)}\ & h>g\rho/2+\sqrt{2t(2t+g\rho)}:\nonumber\\
 &E_1(0)=E_\uparrow(0),\ E_1(\pi)=E_\uparrow(\pi),\  \nu_{\rm IS}=0,\\[2mm]
 {\rm (ii)}\ & g\rho/2-t<h<g\rho/2+\sqrt{2t(2t+g\rho)}:\nonumber\\
&E_1(0)=E_\uparrow(0),\ E_1(\pi)=E_\downarrow(\pi),\  \nu_{\rm IS}=1,\\[2mm]
 {\rm (iii)}\ & h<g\rho/2-2t:\nonumber\\
 &E_1(0)=E_\downarrow(0),\ E_1(\pi)=E_\downarrow(\pi),\ \nu_{\rm IS}=0.
\end{align}
\label{eq:nu_IS_kitaev}
\end{subequations}
In other words, the topological phase transition occurs at $h=g\rho/2+\sqrt{2t(2t+g\rho)}$ and $h=g\rho/2-2t$.
From the above discussion, topological phase diagram is obtained as shown in Fig.~\ref{fig:gs_topo_kitaev} (b),
where the cyan-colored region is topologically non-trivial.
We also calculate the Berry phase by means of the Wilson loop method~\cite{doi:10.1143/JPSJ.74.1674} using the numerically obtained eigenstates of the Bogoliubov equation. Figure~\ref{fig:gs_topo_kitaev} (c) shows the Berry phase a s a function of $h/\lambda$ for $g\rho/\lambda=0.5$ and $t/\lambda=1.5$, which agrees with the analytical result of Eq.~\eqref{eq:nu_IS_kitaev}, i.e., the Berry phase discontinuously changes at $h=g\rho/2+\sqrt{2t(2t+g\rho)}$ and $h=g\rho/2-2t$.
In Fig.~\ref{fig:gs_topo_kitaev} (b), the region where complex eigenvalues appear are highlighted with red, from which we can confirm that the Berry phase and $\nu_{\rm IS}$ are well-defined even in the presence of complex eigenvalues.

\begin{figure}
    \centering
        	\includegraphics[width=8.6cm]{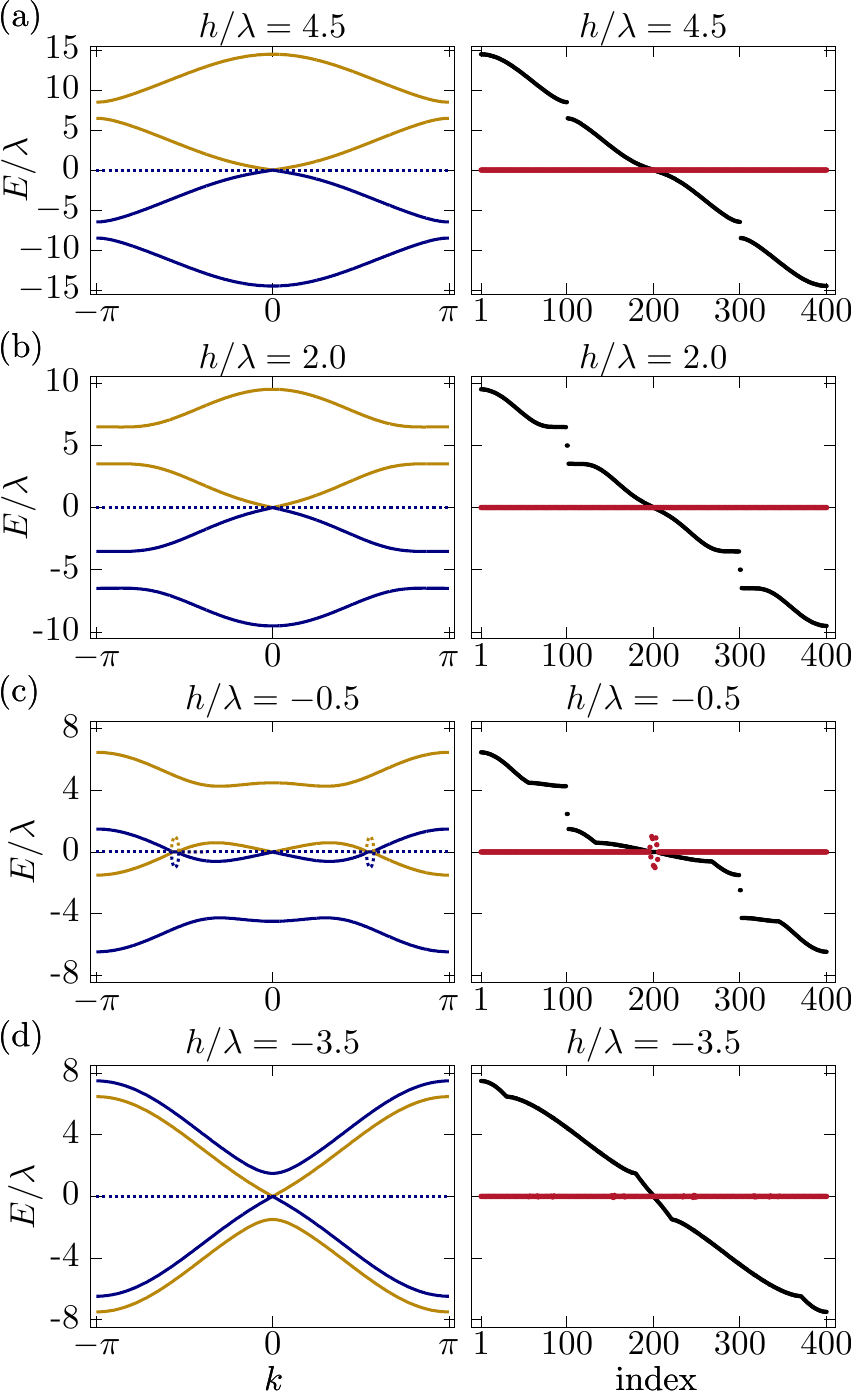}
   	\caption{
   	Excitation spectrum from a BEC in Phase I of the Kitaev-chain-like model calculated for $g\rho/\lambda=0.5$, $t/\lambda=1.5$, and $h/\lambda=4.5$ (a), $2.0$ (b), $-0.5$ (c), and $-3.5$ (d).
	  The left panels are calculated in the momentum space by assuming a periodic boundary condition, where solid and dotted yellow (blue) lines respectively depict ${\rm Re}\,E_n(k)$ and $10\,{\rm Im}\,E_n(k)$ of positive-index (negative-index) bands.
	  The right panels show the eigenvalues obtained in a system of $100$ sites with an open boundary condition, where the black points denote the real parts of the eigenvalues sorted in descending order and the red points are the corresponding imaginary parts multiplied by 10.
	  There are mid-gap states in the right panels of (b) and (c) at $E/\lambda\simeq \pm 5.0$ and $\pm 2.0$, respectively, being consistent with the non-trivial topology: $\nu_{\rm IS}=1$.
	  }
	\label{fig:band_kitaev}
\end{figure}

To see the bulk-edge correspondence, we compare the energy spectra for the systems with periodic and open boundary conditions. Figure~\ref{fig:band_kitaev} shows the energy spectra calculated for $g\rho/\lambda=0.5$, $t/\lambda=1.5$, and several $h/\lambda$. The left panels are calculated in the momentum space by assuming a periodic boundary condition, where solid and dotted yellow (blue) lines respectively depict ${\rm Re}\,E_n(k)$ and $10\,{\rm Im}\,E_n(k)$ of positive-index (negative-index) bands. The right panels show the eigenvalues of the system of $100$ sites with an open boundary condition, where the black points denote the real parts of the eigenvalues sorted in descending order and the red points are the corresponding imaginary parts multiplied by 10.
One can see that there are mid-gap states at $E/\lambda\simeq\pm 5.0$ in the right panel of Fig.~\ref{fig:band_kitaev}~(b) and $E/\lambda\simeq\pm 2.0$ in one of Fig.~\ref{fig:band_kitaev}~(c). We have numerically confirmed from the eigen wavefunction that these states are localized at the edges of the system.
According to the topological phase diagram shown in Fig.~\ref{fig:gs_topo_kitaev}~(b), the topological invariant is $\nu_{\rm IS}=0$ [$\nu_{\rm IS}=1$] for Figs.~\ref{fig:band_kitaev}~(a) and \ref{fig:band_kitaev}~(d) [Figs.~\ref{fig:band_kitaev}~(b) and \ref{fig:band_kitaev}~(c)]. Hence, we can conclude that the nontrivial value of $\nu_{\rm IS}$ is accompanied by edge states. In particular, the bulk-edge correspondence holds even in the presence of complex eigenvalues as seen in the case of Fig.~\ref{fig:band_kitaev}~(c).

Because this model is similar to the Kitaev chain, one might expect a non-trivial topological invariant associated with PHS.
According to the discussions in Sec.~\ref{subsec:topological_invariant_phs}, we can successfully define $\nu_{\rm PHS}$ in the present case:
A typical band structure with complex eigenvalues is that shown in the right panel of Fig.~\ref{fig:band_kitaev}~(c),
for which we can choose an integration path with a fixed imaginary part of the momentum.
(Note that the labeling rule is different from the case of calculating $\nu_{\rm IS}$.)
However, the obtained $\nu_{\rm PHS}$ was always trivial.
\subsection{SSH-like model\label{sec:toy_model_ssh}} 
We consider a 1D spinless bosons in an optical double-well lattice potential~\cite{PhysRevLett.42.1698}. This system is similar to the SSH model, a model that describes electrons in polyacetylene which has TRS, PHS, and hence CS.
There are previous works that discuss non-Hermitian SSH models~\cite{PhysRevA.87.012118,Schomerus:13,PhysRevLett.116.133903,PhysRevB.97.045106,PhysRevB.98.115135,PhysRevLett.121.086803}.
Here we consider the Hamiltonian $\mathcal{\hat{H}}$ given by
\begin{align}\label{Hamiltonian2}
	\mathcal{\hat{H}}
	=
	\sum_{j}\Big[ 
	&-t_1(\hat{a}^{\dagger}_{j,\rm{A}}\hat{a}_{j,\rm{B}} + \hat{a}^{\dagger}_{j,\rm{B}}\hat{a}_{j,\rm{A}})\nonumber\\
	&-t_2(\hat{a}^{\dagger}_{j+1,\rm{A}}\hat{a}_{j,\rm{B}} + \hat{a}^{\dagger}_{j,\rm{B}}\hat{a}_{j+1,\rm{A}})\nonumber\\
	&+\frac{g}{2}(\hat{a}^{\dagger}_{j,\rm{A}}\hat{a}^{\dagger}_{j,\rm{A}}\hat{a}_{j,\rm{A}}\hat{a}_{j,\rm{A}}
	+\hat{a}^{\dagger}_{j,\rm{B}}\hat{a}^{\dagger}_{j,\rm{B}}\hat{a}_{j,\rm{B}}\hat{a}_{j,\rm{B}})\big].
\end{align}
where $\hat{a}_{j,{\rm S}}^\dagger$ ($\hat{a}_{j,{\rm S}}$) is the creation (annihilation) operator of an atom at the sublattice S=A or B in the $j$th cite, $t_1 (t_2)$ is the intra-cell (inter-cell) hopping amplitude, and $g$ is the inter-atomic interaction.
The parameters $t_1, t_2$, and $g$ are all real, and hence $\hat{H}$ is an Hermitian operator.
\begin{figure}
	\centering
		\includegraphics[width=6.0cm]{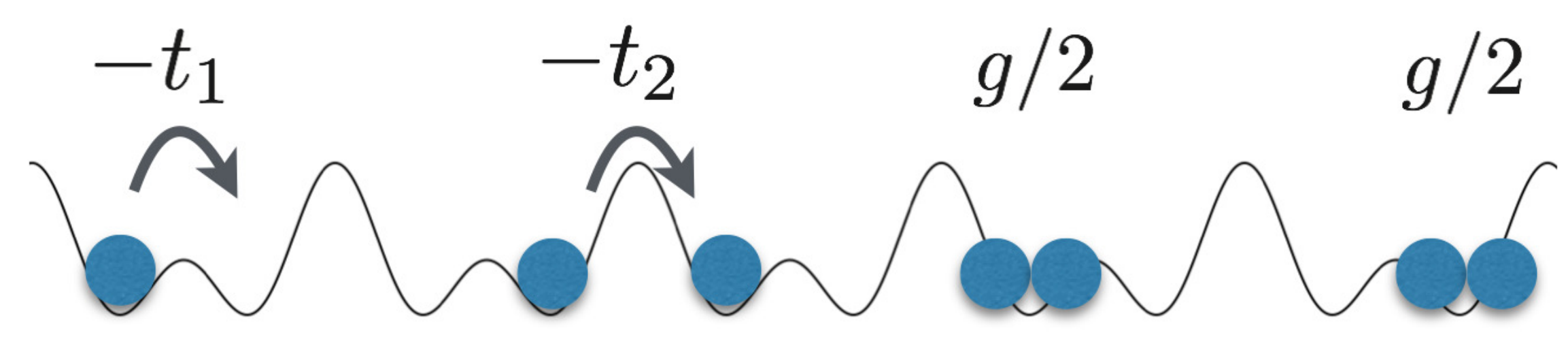}
	\caption{
	Schematic of the SSH-like model given by Eq.~\eqref{Hamiltonian2}.
	Spinless bosons are loaded into a 1D optical double-well lattice potential.
	Here, $t_1$ ($t_2$) is the intra-cell (inter-cell) hopping amplitude and 
	$g$ is the inter-atomic interaction.
	}
	\label{ssh_model}
\end{figure}

\begin{figure}
    \centering
        	\includegraphics[width=8.6cm]{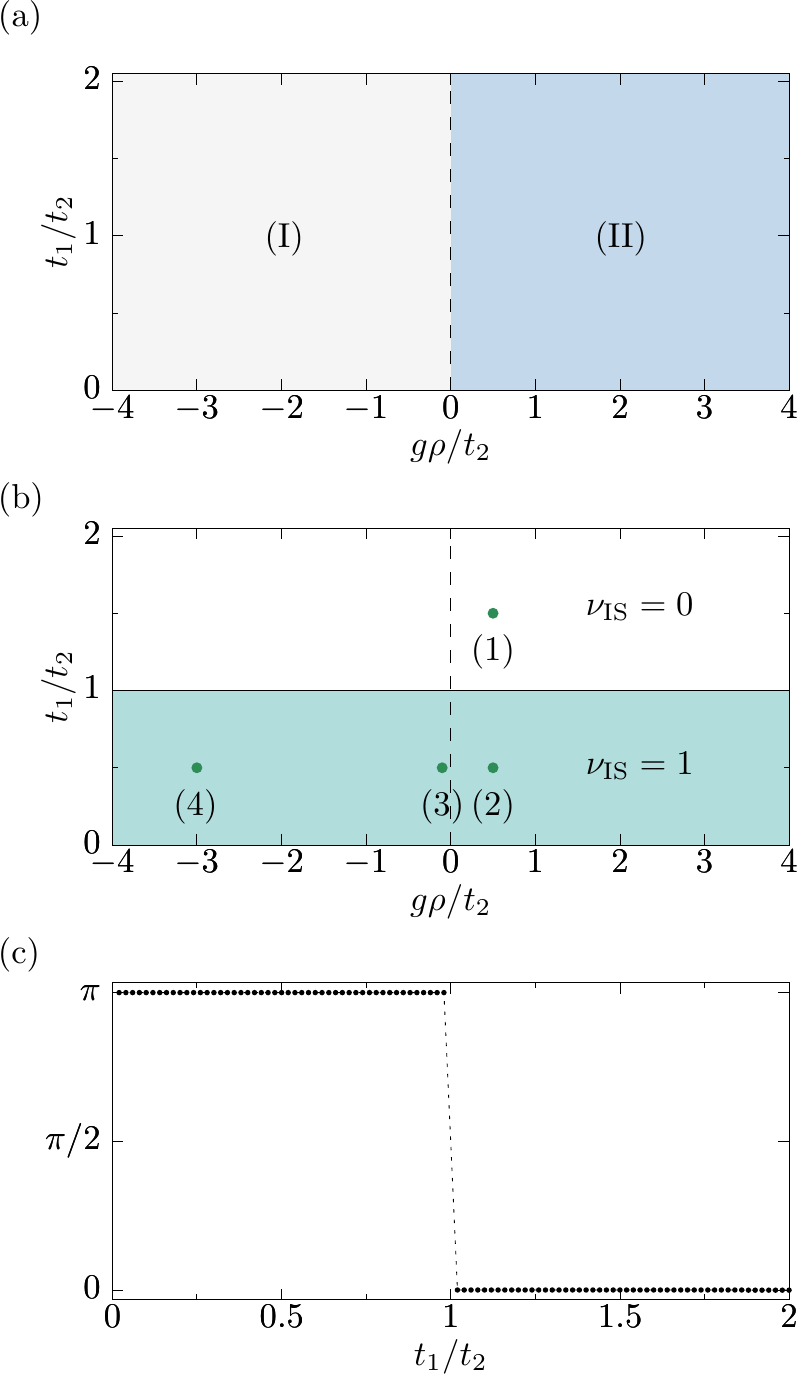}
   	\caption
	{
	(a) Ground-state phase diagram of the SSH-like model given by Eq.~\eqref{Hamiltonian2}.
	There are two phases. In Phase I, atoms are condensed into one sublattice at a certain cite due to the attractive interaction.
	In Phase II, atoms are uniformly distributed in both sublattices.
	We prepare a BEC in Phase II and calculate the Bogoliubov bands with changing $t_1/t_2$ and $g\rho/t_2$.
	Complex eigenvalues appear for $g\rho/t_2<0$.
	(b) Topological phase diagram for the Bogoliubov bands from a BEC in Phase II. The dashed lines depict the phase boundary shown in (a).
	Topological phase transition points are given by $t_1/t_2=1$,
	and the $\mathbb{Z}_2$ topological invariant $\nu_{\rm IS}$ becomes non-trivial in the cyan-colored region. 
	The points labeled with (1)-(4) indicate the parameters used in the calculation of Figs. \ref{fig:band_ssh}~(a)-\ref{fig:band_ssh}~(d), respectively.
	(c) Berry phase as a function of $t_1/t_2$ numerically calculated by means of the Wilson loop method.
	The result does not depend on the value of $g\rho/t_2$.
	The Berry phase discontinuously jumps between $0$ and $\pi$ at $t_1/t_2=1$, being in good agreement with the phase transition line shown in (b).
	}
    \label{fig:gs_topo_ssh}
\end{figure}
Assuming a superfluid phase,
we obtain the ground-state phase diagram as shown in Fig.~\ref{fig:gs_topo_ssh} (a).
There are only two phases.
When the interaction is attractive, i.e., $g<0$, all atoms are condensed into one of the sublattices at a certain cite. This is Phase I shown in Fig.~\ref{fig:gs_topo_ssh} (a).
On the other hand, when the interaction is repulsive ($g>0$), atoms are uniformly distributed in both sublattices. This is Phase II in Fig.~\ref{fig:gs_topo_ssh} (a). The order parameter for Phase II is given by $(\langle \hat{a}_{j,{\rm A}}\rangle,\langle \hat{a}_{j,{\rm B}}\rangle)=\sqrt{\rho}(1,1)$ where $\rho=N/(2L)$ is the mean number of atoms per sublattice with $N$ being the total number of atoms and $L$ the number of double well.

Below, we prepare a BEC in Phase II and investigate the excitation spectrum with changing $t_1/t_2$ and $g/t_2$. 
We perform a Fourier transform $\hat{a}_{j,\rm{S}}=(1/\sqrt{L})\sum_{k}\hat{a}_{k,\rm{S}}e^{ikj}$ with taking the lattice constant to be unity and obtain the Bogoliubov equation:
\begin{subequations}
\begin{align}
	H(k)T(k)&=
	T(k)
	\begin{pmatrix}
		E(k) & 0\\
		0 & -E^*(k)
	\end{pmatrix}
	,\\
	H(k)&=
	\begin{pmatrix}
		H^{(1)}(k) & H^{(2)}\\
		-[H^{(2)}]^* & -[H^{(1)}(-k)]^*
	\end{pmatrix}
	,
	\\
	H^{(1)}(k)&=
	\begin{pmatrix}
		-\mu + 2g\rho & -t_1-t_2e^{-ik}\\
		-t_1-t_2e^{ik} & -\mu + 2g\rho
	\end{pmatrix}
	,
	\\
	H^{(2)}&=
	\begin{pmatrix}
		g\rho & 0\\
		0 & g\rho
	\end{pmatrix}
	.
\end{align}
\end{subequations}
where $\mu=-t_1-t_2+g\rho$ is the chemical potential for the prepared state.

The Bogoliubov equation for the SSH-like model preserves TS and IS, and hence $H(k)$ satisfies the following relations:
\begin{align}
    \begin{array}{rrlrl}
	\textrm{pseudo-H:}& \eta H(k){\eta}^{-1}&=\ \ {H(k)}^{\dagger},&\eta&=\tau_3,\\[1.5mm]
	{\rm PHS:}& \mathcal{C} H(k)\mathcal{C}^{-1}&=-H(-k),\ &\mathcal{C}&=\tau_1K,\\[1.5mm]
	{\rm TS:}& \Theta H(k)\Theta^{-1}&=\ \ H(-k),\ &\Theta&=K,\\[1.5mm]
	{\rm CS:}& \Gamma H(k)\Gamma^{-1}&=-H(k),\ &\Gamma&=\tau_1,\\[1.5mm]
	{\rm IS:}& \mathcal{P} H(k)\mathcal{P}^{-1}&=\ \ H(-k),\ &\mathcal{P}&=\sigma_1,
    \end{array}
\end{align}
where $\tau_{i=0,1,2,3}$ ($\sigma_{i=0,1,2,3}$) is the Pauli matrices in the Nambu (sublattice) space.

We first calculate $\nu_{\rm IS}$ with $n_{\rm r}=1$. Since $H(k_{\rm IS}=0,\pi)$ commutes with the inversion operator $\mathcal{P}=\sigma_1$,
$H(k_{\rm IS})$ is divided into two sectors of states with $\mathcal{P}$ eigenvalues $+1$ and $-1$. To be more concrete, defining a unitary matrix $U=(\sigma_0-i\sigma_2)/\sqrt{2}$ that satisfies $U^\dagger \sigma_1 U =\sigma_3$, $H(k_{\rm IS})$ is transformed to
\begin{widetext}
\begin{align}
    &U^\dagger H(k_{\rm IS}) U =\nonumber\\
    &\begin{pmatrix}
    -\mu+2g\rho-t_1-t_2\cos k_{\rm IS} & 0 & g\rho & 0\\
    0 & -\mu+2g\rho+t_1+t_2\cos k_{\rm IS}& 0 & g\rho\\
    -g\rho & 0 & \mu-2g\rho+t_1+t_2\cos k_{\rm IS} & 0 \\
    0 & -g\rho & 0 & \,\mu-2g\rho-t_1-t_2\cos k_{\rm IS}
\end{pmatrix}.
\end{align}
\end{widetext}
Then, the eigenvalues of $H(k_{\rm IS})$ is obtained by diagonalizing the $2\times2$ matrices, resulting in
\begin{subequations}
\begin{align}
    E_\uparrow(0) &= 0,\\
    E_\downarrow(0) &= 2\sqrt{(t_1+t_2)(t_1+t_2+g\rho)},\\
    E_\uparrow(\pi) &= 2\sqrt{t_2(t_2+g\rho)},\\
    E_\downarrow(\pi) &= 2\sqrt{t_1(t_1+g\rho)},
\end{align}
\end{subequations}
where the subscript $\uparrow$ ($\downarrow$) indicates that the $\mathcal{P}$ eigenvalue of the corresponding eigenstate is $+1$ ($-1$).
Here, we assume $t_1,t_2>0$. Then, the two bands touch at $k=0$ at $t_1+t_2+g\rho=0$, whereas they touch at $k=\pi$ at $t_1=t_2$ and $t_1+t_2+g\rho=0$.
As a result, the band inversion occurs at $t_1=t_2$ independently of the value of $g\rho$.
(At $t_1+t_2+g\rho=0$, $\xi_1(0)$ and $\xi_1(\pi)$ change simultaneously.)
Considering the band structure in the whole BZ, we obtain $\nu_{\rm IS}=0$ for $t_1>t_2$ and $\nu_{\rm IS}=1$ for $t_1<t_2$.
The resulting topological phase diagarm is shown in Fig.~\ref{fig:gs_topo_ssh} (b).
We also numerically calculate the Berry phase by means of the Wilson loop method.
The result is shown in Fig.~\ref{fig:gs_topo_ssh} (c), from which we can confirm that the topologically non-trivial phase arises at $t_1/t_2<1$.

\begin{figure}[htb]
    \centering
		\includegraphics[width=8.6cm]{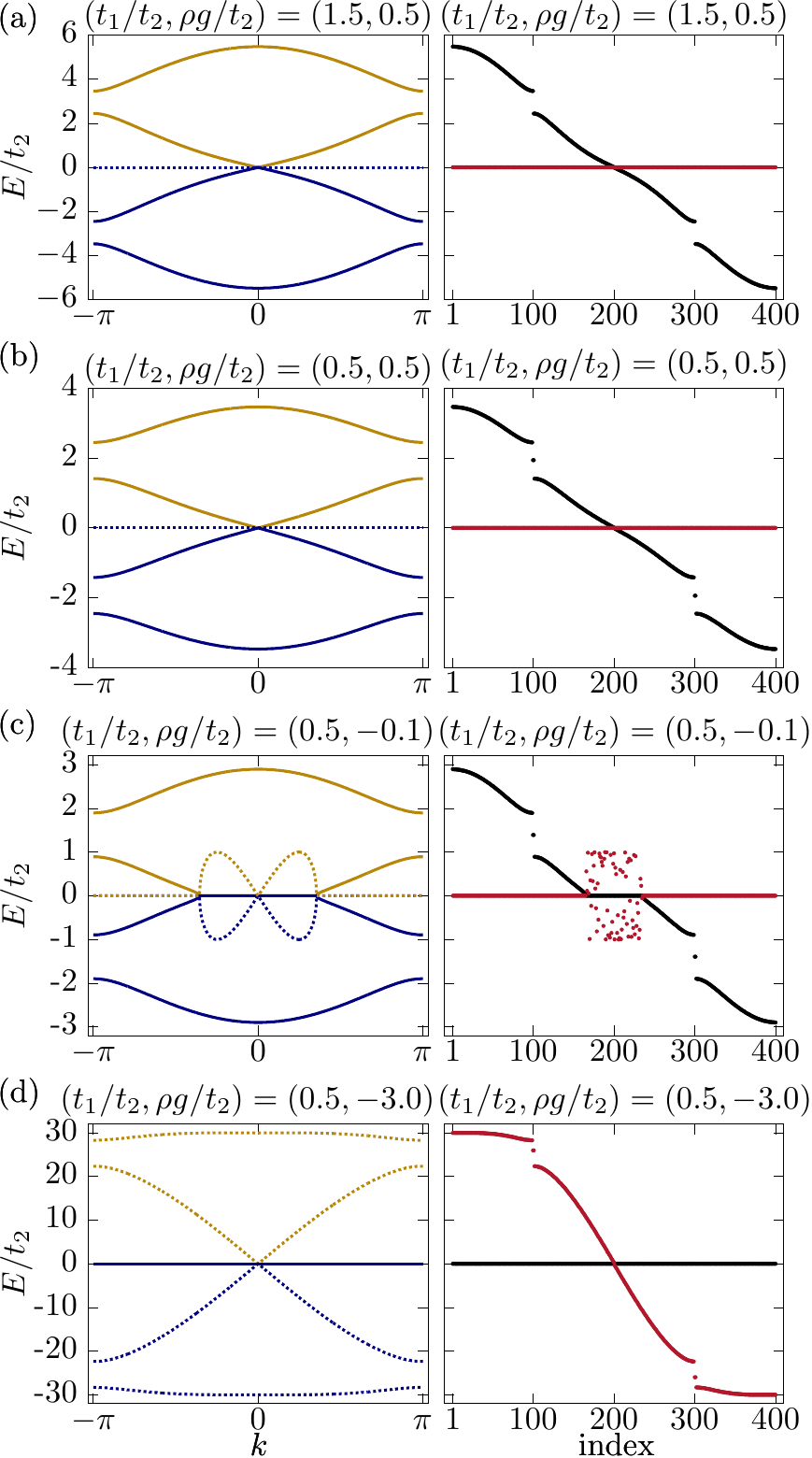}
   	  \caption
	  {
	  Excitation spectrum from a BEC in Phase II of the SSH-like model calculated for $(t_1/t_2, g\rho/t_2)=(1.5,0.5)$ (a),
	   $(0.5,0.5)$ (b), $(0.5,-0.1)$ (c), and $(0.5,-3.0)$ (d).
	  These parameter sets are indicated in Fig.~\ref{fig:gs_topo_ssh}~(a) by points labeled with (1)-(4), respectively.
    The meaning of the panels and lines in them are the same as those in Fig.~\ref{fig:band_kitaev}.
	  There are mid-gap states in the right panels of (b), (c), and (d) at $E/t_2\simeq \pm 1.9$, $\pm1.4$, and $\pm 26$, respectively, being consistent with the non-trivial topology: $\nu_{\rm IS}=1$.
     }
	\label{fig:band_ssh}
\end{figure}

To see the bulk-edge correspondence, we compare the energy spectra for the systems with periodic and open boundary conditions.
Figure~\ref{fig:band_ssh} shows the results for various values of $t_1/t_2$ and $g\rho/t_2$,
where the meaning of the lines are the same as those in Fig.~\ref{fig:band_kitaev}.
The values of $(t_1/t_2,g\rho/t_2)$ used in Fig.~\ref{fig:band_ssh} (a)-\ref{fig:band_ssh}(d) are indicated in the phase diagram of Fig.~\ref{fig:gs_topo_ssh}~(a) by points labeled with (1)-(4), respectively.
In accordance with $\nu_{\rm IS}=1$, mid-gap states exist in the right panel of Figs.~\ref{fig:band_ssh}~(b), \ref{fig:band_ssh}~(c), and \ref{fig:band_ssh}~(d).
In particular in Fig.~\ref{fig:band_ssh}~(d), although all eigenvalues are complex in the whole BZ, there exist edge states that has pure imaginary eigenvalue.

Next, we show that the winding number defined in Eq.~\eqref{def:winding_number} is zero. In the present case, the Hermitian Hamiltonian Eq.~\eqref{Hamiltonian_0} is given by
\begin{align}
    H_0&=\begin{pmatrix}H^{(1)}(k) & 0 \\ 0 & -H^{(1)}(k) \end{pmatrix} =H^{(1)}(k)\tau_3,
\end{align}
where we have used the fact that $[H^{(1)}(-k)]^*=H^{(1)}(k)=[H^{(1)}(k)]^\dagger$ and $[H^{(2)}]^{\rm T}=H^{(2)}$. 
The chiral operator $\tilde{\Gamma}$ that anti-commutes with $H_0(k)$ is given by
\begin{align}
	\tGamma=i\tau_3\Gamma=-\tau_2.
\end{align}
As we mentioned in Sec.~\ref{sec:winding_number_CS}, this $\tGamma$ anti-commutes with $\tau_3$:
\begin{align}
\label{anti_com_tGamma_tau_3}
	\{\tGamma,\tau_3\}=0.
\end{align}
The unitary matrix $U_{\tGamma}$ that diagonalizes $\tGamma$ is given by
\begin{align}
    U_{\tGamma}=\frac{1}{\sqrt{2}}(\tau_0-i\tau_1),
\end{align}
with which $H_0(k)$ is transformed to
\begin{align}
    U_{\tGamma}^\dagger H_0(k) U_{\tGamma}=H^{(1)}(k)\tau_2=\begin{pmatrix}0 & -i H^{(1)}(k) \\ iH^{(1)}(k) & 0\end{pmatrix}.
\end{align}
It follows that $\tilde{q}(k)=-iH^{(1)}(k)$ and ${\rm det}[\tilde{q}(k)] = -{\rm det}[H^{(1)}(k)]$.
Since $H^{(1)}(k)$ is an Hermitian matrix, ${\rm det}[\tilde{q}(k)]$ is always real and thus, the winding number given by Eq.~\eqref{def:winding_number} is zero as we predicted.

We have also calculated $\nu_{\rm PHS}$. For the case of band structures shown in Fig.~\ref{fig:band_ssh}~(a)-\ref{fig:band_ssh}~(c),
we can successfully define the integration path and hence $\nu_{\rm PHS}$, which is however shown to be trivial.
For the case of band structure shown in fig.~\ref{fig:band_ssh}~(d), $\nu_{\rm PHS}$ cannot be defined
because complex eigenvalues appear at $k=0$ and $\pi$.

\section{conclusion\label{sec:conclusion}}
In this work, we have studied the topology of bosonic Bogoliubov bands from a BEC trapped in a 1D optical lattice, in particular, in the presence of complex eigenvalues.
Contrarily to the fermionic ones, the bosonic Bogoliubov systems are non-Hermitian, and hence complex eigenvalues arise in various cases,
known as dynamical instability.
The appearance of complex eigenvalues, in general, accompanies EPs, at which the Berry connection is ill-defined.
We have therefore discussed how to define topological invariants in the presence of EPs.
We have found that EPs can be removed from the BZ by introducing an imaginary part of the momentum.
In other words, we can avoid EPs by taking an integration path on the complex plane and define topological invariants which are a simple
generalization of those defined in Hermitian systems.
Topological invariants associated with IS and PHS are defined by the Berry phase, for which the integration path on the complex plane is
dependent on the symmetry we are focusing on.
We have also shown that the winding number associated with CS can be defined regardless of the existence of EPs but that it is always
trivial.
To see the bulk-edge correspondence, we have numerically investigated two toy models with IS,
and confirmed that edge states appear when the $\mathbb{Z}_2$ topological invariant $\nu_{\rm IS}$ is nontrivial.
In particular, in Fig.~\ref{fig:band_ssh}~(d) the edge states are clearly identified even when bulk eigenvalues are all complex.
We have also numerically calculated the Berry phase by means of the Wilson loop method, which agrees with the analytically obtained $\nu_{\rm IS}$.
This result also suggests that the Berry phase is well-defined even in the presence of complex eigenvalues.

There are several remaining issues.
First, though we have defined the Berry phase of the hole bands as a topological invariant associated with PHS,
it is shown to belong to $\mathbb{Z}_2$ only for special band structures.
One may wonder which topological class our system belongs to and what the topological invariant that characterizes the system should be.
These are still open questions.
Classification of non-Hermitian Hamiltonians has been done in Ref.~\cite{1812.09133},
which is however applicable only for gapful systems.
To apply their results to the present case,
the distributions of the eigenvalues of particle and hole bands should be separated in complex energy space,
which means that the band structures shown in Figs.~\ref{fig:band_kitaev}~(c) and \ref{fig:band_ssh}~(c) are out of their classification.
The second issue is what the topological charge that characterizes EPs is.
The topological charge of EPs is discussed in several papers~\cite{PhysRevLett.120.146402, PhysRevA.72.014104, PhysRevA.87.012118, PhysRevLett.118.045701}.
Using the notation in Ref.~\cite{PhysRevLett.120.146402}, all EPs in the present system has vorticity $\pm1/2$.
In addition to this, the Berry phase along a contour enclosing an EP in complex momentum space might be quantized, 
which will be discussed in future work.
Extension to a two-dimensional system is also interesting.
As we have shown in Sec.~\ref{sec:Mathematical framework relating to an exceptional point}, 
EPs in a two-dimensional system appear as a ring.
It might be possible to define the Chern number in the presence of an exceptional ring
by introducing an imaginary part of the two-dimensional momentum.

\begin{acknowledgments}
The authors would like to thank A. Yamakage, S. Tamura, K. Kawabata, K. Mochizuki, and H. Obuse for useful discussions. 
This work was supported by JST-CREST (Grant No. JPMJCR16F2) and JSPS KAKENHI
Grant Nos. JP15K17726 and 19H01824.
S.K. was supported by JSPS KAKENHI Grant Nos. JP17H02922 and JP19K14612, and the Building of Consortia for the Development of Human Resources in Science and Technology.
\end{acknowledgments}

\appendix
\section{Emergence of an exceptional point in a pseudo-Hermitian system}\label{sec:appendix_EP}
We show that an EP appears where two real eigenvalues change into a complex conjugate pair.
In the vicinity of such a point, the band behavior is effectively described by a $2\times 2$ matrix corresponding to the two coalescing elements:
\begin{align}
H(k)=a_0'(k)\sigma_0+a_1'(k)\sigma_1+a_2'(k)\sigma_2+a_3'(k)\sigma_3,
\label{eq:2x2H}
\end{align}
where $a_i'(k)\ (i=0,1,2,3)$ are complex functions of $k$, $\sigma_0$ is the $2\times 2$ identity matrix, and $\sigma_{1,2,3}$ are the Pauli matrices.
Imposing the pseudo-Hermiticity $\sigma_3H(k)\sigma_3=H^\dagger(k)$, we obtain that $a_0'(k)$ and $a_3'(k)$ ($a_1'(k)$ and $a_2'(k)$) are real (pure imaginary), i.e., $H(k)$ is rewritten as
\begin{align}
H(k)&=a_0(k)\sigma_0+ia_1(k)\sigma_1+ia_2(k)\sigma_2+a_3(k)\sigma_3\nonumber\\
&=\begin{pmatrix} a_0(k)+a_3(k) & ia_1(k)+a_2(k) \\ ia_1(k)-a_2(k) & a_0(k)-a_3(k)\end{pmatrix},
\end{align}
with $a_i(k)$ beging real functions of $k$.
The eigenvalues of $H(k)$ are given by
\begin{align}
E_{\pm}(k) &= a_0(k) \pm \sqrt{E_{\rm root}(k)}\\
E_{\rm{root}}(k) &:=  a^2_3(k) - a^2_1(k) - a^2_2(k)
\end{align}
Thus, the bands touch at the point where $E_{\rm root}(k)=0$.
The two eigenvalues are both real in the one side of the band touching point,
whereas they are complex conjugate to each other in the other side.
When we expand $E_{\rm root}(k)$ around its zero point, say $k=k_0$, the leading order term should be linear in $k-k_0$, 
resulting in the square root behavior of the eigenvalues as discussed in the main text.
The unnormalized right and left eigenstates belonging to $E_\pm(k)$ are respectively given by
\begin{align}
	\ket{\ur_\pm(k)} &= \begin{pmatrix} a_3\pm \sqrt{E_{\rm root}(k)} \\  ia_1+a_2\end{pmatrix},\label{eq:urpm}\\
	\ket{\ul_\pm(k)} &= \begin{pmatrix} a_3\pm (\sqrt{E_{\rm root}(k)})^* \\ -ia_1-a_2\end{pmatrix}.\label{eq:ulpm}
\end{align}
One can see from Eqs.~\eqref{eq:urpm} and \eqref{eq:ulpm} that $\ket{u^{\rm R/L}_+(k)}$ and $\ket{u^{\rm R/L}_-(k)}$ are identical to each other at $E_{\rm root}(k)=0$.
Instead, they satisfy the self-orthogonality at this point:
\begin{align}
	\braket{\ul_\pm(k)|\ur_\pm(k)} &=  \left.2 \sqrt{E_{\rm{root}}(k)}(\sqrt{E_{\rm{root}}(k)} \pm a_3)\right|_{E_{\rm root}(k)=0}\nonumber\\
&=0.
\end{align}
Hence, $E_{\rm root}(k)=0$ is an EP.

We note that for the case of a bosonic Bogoliubov system the above discussion is valid only when the positive-norm and negative-norm bands (i.e., particle and hole bands) touch.
Since the metric operator of the pseudo-Hermiticity is a matrix in the Nambu space, the effective $2\times 2$ Hamiltonian \eqref{eq:2x2H}
satisfies the pseudo-Hermiticity relation only when it is constituted by one band from the particle sector and another band from the hole sector.
When two positive-norm bands or two negative-norm bands touch, the effective Hamiltonian becomes Hermitian rather than pseudo-Hermitian.
To summarize, an EP in a bosonic Bogoliubov system appears at a point where positive-norm and negative-norm bands touch and change into a pair of complex-eigenenergy bands.


\section{Proof of Eq.~\eqref{eq:extended_sym}}
\label{sec:appendix_extended_sym}
We expand the each Nambu sector of $H(\kappa)$ defined in Eq.~\eqref{eq:def_H_kappa} with respect to $\kappa$ as
\begin{align}
H^{(1)}(\kappa)=\sum_n c_n \kappa^n,\ \ \ 
H^{(2)}(\kappa)=\sum_n d_n \kappa^n,
\end{align}
where $c_n$ and $d_n$ are $\mathcal{N}\times\mathcal{N}$ matrices.
Then the whole matrix $H(\kappa)$ is expanded as
\begin{align}
&H(\kappa)\nonumber\\
&=
\sum_{n:{\rm even}} \kappa^n \begin{pmatrix}
c_n & d_n \\ -d_n^* & -c_n^* \end{pmatrix}
+\sum_{n:{\rm odd}} \kappa^n \begin{pmatrix}
c_n & d_n \\ d_n^* & c_n^* \end{pmatrix}.
\label{eq:H_kappa1}
\end{align}

{\it Particle-hole Symmetry}:
Operating $\mathcal{C}=\mathcal{C}^{-1}=\tau_1 K$ from the both side of Eq.~\eqref{eq:H_kappa1}, we obtain
\begin{align}
&\mathcal{C}H(\kappa)\mathcal{C}^{-1}\nonumber\\
&=
\sum_{n:{\rm even}} (\kappa^*)^n \tau_1\begin{pmatrix}
c_n^* & d_n^* \\ -d_n & -c_n \end{pmatrix}\tau_1
+\sum_{n:{\rm odd}} (\kappa^*)^n \tau_1\begin{pmatrix}
c_n^* & d_n^* \\ d_n & c_n \end{pmatrix}\tau_1\nonumber\\
&=
\sum_{n:{\rm even}} (\kappa^*)^n \begin{pmatrix}
-c_n & -d_n \\ d_n^* & c_n^* \end{pmatrix}
+\sum_{n:{\rm odd}} (\kappa^*)^n \begin{pmatrix}
c_n & d_n \\ d_n^* & c_n^* \end{pmatrix}\nonumber\\
&= -H(-\kappa^*).
\label{eq:extendedPHS_appendix}
\end{align}

{\it Time-reversal Symmetry}:
When $H(k)$ preserves TRS for real $k$, substituting Eq.~\eqref{eq:H_kappa1} with $\kappa=k\in \mathbb{R}$ to Eq.~\eqref{eq:TRS} results in
\begin{align}
    Uc_n^*U^{-1}&=(-1)^nc_n, \\
    U d_n^* U&=(-1)^nd_n,
\end{align}
where we have used Eq.~\eqref{eq:Theta}.
Then, operating $\Theta$ and $\Theta^{-1}$ from the both side of Eq.~\eqref{eq:H_kappa1}, we obtain
\begin{align}
    &\Theta H(\kappa)\Theta^{-1}\nonumber\\
    &=
\sum_{n:{\rm even}} (\kappa^*)^n \begin{pmatrix}
c_n & d_n \\ -d_n^* & -c_n^* \end{pmatrix}
- \sum_{n:{\rm odd}} (\kappa^*)^n \begin{pmatrix}
c_n & d_n \\ d_n^* & c_n^* \end{pmatrix}\nonumber\\
&=H(-\kappa^*).
\label{eq:extendedTS_appendix}
\end{align}

{\it Chiral Symmetry}:
From Eqs.~\eqref{eq:extendedPHS_appendix} and \eqref{eq:extendedTS_appendix}, we obtain
\begin{align}
    \Gamma H(\kappa) \Gamma^{-1} = \mathcal{C} H(-\kappa^*) \mathcal{C}^{-1} = -H(\kappa).
\end{align}

{\it Inversion Symmetry}:
When $H(k)$ preserves IS for real $k$, substituting Eq.~\eqref{eq:H_kappa1} with $\kappa=k\in \mathbb{R}$ to Eq.~\eqref{eq:IS} results in
\begin{align}
    \mathcal{P}\begin{pmatrix}
c_n & d_n \\ -d_n^* & -c_n^* \end{pmatrix}\mathcal{P}^{-1}
&=\begin{pmatrix}
c_n & d_n \\ -d_n^* & -c_n^* \end{pmatrix}\\
\mathcal{P} \begin{pmatrix}
c_n & d_n \\ d_n^* & c_n^* \end{pmatrix}\mathcal{P}^{-1}
&= - \begin{pmatrix}
c_n & d_n \\ d_n^* & c_n^* \end{pmatrix}.
\end{align}
Then, operating $\mathcal{P}$ and $\mathcal{P}^{-1}$ from the both side of Eq.~\eqref{eq:H_kappa1}, we obtain
\begin{align}
    &\mathcal{P}H(\kappa)\mathcal{P}^{-1}\nonumber\\
    &=\sum_{n:{\rm even}} \kappa^n \begin{pmatrix}
c_n & d_n \\ -d_n^* & -c_n^* \end{pmatrix}
-\sum_{n:{\rm odd}} \kappa^n \begin{pmatrix}
c_n & d_n \\ d_n^* & c_n^* \end{pmatrix}\nonumber\\
    &=H(-\kappa).
\end{align}

\section{Proof of of Eq.~\eqref{sum_rule}}\label{sec:appendix_prof_sum_rule}
Using the Bogoliubov transformation matrix
\begin{align}
T^{\rm R}(k):=\begin{pmatrix} |u^{\rm R}_{-\mathcal{N}}(k)\rangle & |u^{\rm R}_{-\mathcal{N}+1}(k)\rangle & \cdots& |u^{\rm R}_{\mathcal{N}}(k)\rangle\end{pmatrix}
\end{align}
and its inverse matrix
\begin{align}
[T^{\rm L}(k)]^\dagger:=\begin{pmatrix}\langle u^{\rm L}_{-\mathcal{N}}(k)| \\ \langle u^{\rm L}_{-\mathcal{N}+1}(k)| \\ \vdots \\ \langle u^{\rm L}_{\mathcal{N}}(k) |\end{pmatrix}=[T^{\rm R}(k)]^{-1},
\end{align}
the  matrix is rewritten as
\begin{align}
\mathcal{A}(k)
&=\frac{i}{2}\left\{ [T^{\rm L}(k)]^\dagger \partial_k T^{\rm R}(k) + [T^{\rm R}(k)]^\dagger \partial_k T^{\rm L}(k)\right\}\nonumber\\
&=\frac{i}{2}\left\{ [T^{\rm R}(k)]^{-1} \partial_k T^{\rm R}(k) + [T^{\rm L}(k)]^{-1} \partial_k T^{\rm L}(k)\right\}\nonumber\\
&=\frac{i}{2}\left\{ \partial_k \log T^{\rm R}(k) + \partial_k \log T^{\rm L}(k)\right\},
\end{align}
from which the left-hand side of the Eq.~\eqref{berry_all} is calculated as
\begin{align}
	\int^{\pi}_{-\pi}\frac{dk}{2\pi}A(k)
	&= \frac{i}{2} \int^{\pi}_{-\pi}\frac{dk}{2\pi} \partial_{k}\left\{ \log\det[T^{\rm{R}}(k)T^{\rm{L}}(k)]\right\}.
\label{eq:append_Ak}
\end{align}
From the relationship $[T^{\rm L}(k)]^\dagger T^{\rm R}(k)={\bf 1}$, we obtain $\mbox{det}[T^{\rm L}(k)]=1/\mbox{det}[T^{\rm{R}}(k)]^*$,
and the right-hand side of Eq.~\eqref{eq:append_Ak} is given by
\begin{align}
	\int^{\pi}_{-\pi}\frac{dk}{2\pi}A(k)
 	 &=\frac{i}{2} \int^{\pi}_{-\pi}\frac{dk}{2\pi} \partial_k \Big[ \log e^{2i\theta(k)}\Big]\nonumber\\
 	 &=-\frac{\theta(\pi)-\theta(-\pi)}{2\pi}
	 ,
\end{align}
where $\theta(k):={\rm arg}\,{\rm det}[T^{\rm R}(k)]$. Since $T^{\rm{R}}(k)$ is a $2\pi$ periodic function of $k$, $\theta(\pi)-\theta(-\pi)$ is integer multiples of $2\pi$, i.e.,
\begin{align}
	\int^{\pi}_{-\pi}\frac{dk}{2\pi}A(k) =N\in \mathbb{Z} \nonumber.
\end{align}

\section{Proof of of Eq.~\eqref{eq:A(k)-A(-k)-2}\label{sec:Append_IS}}
We start from the inversion symmetry for a complex momentum $\kappa$:
\begin{align}
	\mathcal{P}H(\kappa)\mathcal{P}^{-1}&=H(-\kappa).
\end{align}
Suppose that the $n_{\rm r}$th band and $n_{\rm r}+1$th band are separated. 
Then, when $|u_n^{\rm R,L}(\kappa)\rangle$ with $1\le n\le n_{\rm r}$ is an eigenstate of $H(\kappa)$,
$\mathcal{P}|u_n^{\rm R,L}(\kappa)\rangle$ can be expanded in terms of  $|u_m^{\rm R,L}(-\kappa)\rangle$ with $1\le m\le n_{\rm r}$ as
\begin{subequations}
	\begin{align}\label{inversion}
		\mathcal{P}\ket{u^{\rm{R}}_n(\kappa)}&=\sum_{m=1}^{n_{\rm r}}[\Up^{\rm{LR}}(-\kappa)]_{mn}\ket{u^{\rm{R}}_m(-\kappa)},\\
		\mathcal{P}\ket{u^{\rm{L}}_n(\kappa)}&=\sum_{m=1}^{n_{\rm r}}[\Up^{\rm{RL}}(-\kappa)]_{mn}\ket{u^{\rm{L}}_m(-\kappa)},
	\end{align}
\end{subequations}
where 
\begin{subequations}
\begin{align}
[\Up^{\rm{LR}}(-\kappa)]_{mn}&:= \braket{u^{\rm{L}}_m(-\kappa)|\mathcal{P}|u^{\rm{R}}_n(\kappa)},\\
[\Up^{\rm{RL}}(-\kappa)]_{mn}&:= \braket{u^{\rm{R}}_m(-\kappa)|\mathcal{P}|u^{\rm{L}}_n(\kappa)}
\end{align}
\end{subequations}
are $n_{\rm r}\times n_{\rm r}$ matrices and satisfy $\Up^{\rm{RL}}(\kappa) \Up^{\rm{LR}}(\kappa)^{\dagger}=\bm{1}$. 
From the above relations, we can relate $[\mathcal{A}^{\rm LR}(-\kappa)]_{nn'}$ ($1\le n,n' \le n_{\rm r}$) with $[\mathcal{A}^{\rm LR}(\kappa)]_{mm'}$ ($1\le m,m'\le n_{\rm r}$) as follows:
\begin{align}
	&[\mathcal{A}^{\rm{LR}}(-\kappa)]_{nn'}\nonumber\\
	&=-i\braket{\ul_{n}(-\kappa)|\partial_{\kappa} \ur_{n'}(-\kappa)}
	\nonumber\\
	&=-i\sum_{m,\ell=1}^{n_{\rm r}}[\Up^{\rm{RL}}(\kappa)]^\dagger_{nm}\langle \mathcal{P}\ul_{m}(\kappa)|\partial_{\kappa}\Big([\Up^{\rm{LR}}(\kappa)]_{\ell n'}|\mathcal{P}\ur_{\ell}(\kappa)\rangle\Big)
	\nonumber\\
	&=-i\sum_{m,\ell=1}^{n_{\rm r}}{[\Up^{\rm{RL}}(\kappa)]^\dagger_{nm}}\braket{\ul_{m}(\kappa)| \partial_{\kappa} \ur_{\ell}(\kappa)} {[\Up^{\rm{LR}}(\kappa)]_{\ell n'}}\nonumber\\
	&\ \ \ -i\sum_{m,\ell>0} {[\Up^{\rm{RL}}(\kappa)]^\dagger_{nm}} \partial_{\kappa} [\Up^{\rm{LR}}(\kappa)]_{\ell n'}\delta_{m\ell}
	\nonumber\\
	&=-\sum_{m,\ell=1}^{n_{\rm r}} [\Up^{\rm{RL}}(\kappa)]^\dagger_{nm} [\mathcal{A}^{\rm{LR}}(\kappa)]_{m\ell} [\Up^{\rm{LR}}(\kappa)]_{\ell n'}\nonumber\\
	&\ \ \ -i\sum_{m=1}^{n_{\rm r}} {[\Up^{\rm{RL}}(\kappa)]^\dagger_{nm}} \partial_{\kappa} [\Up^{\rm{LR}}(\kappa)]_{m n'},
\end{align}
from which we obtain
\begin{align}
	&\sum_{n=1}^{n_{\rm r}}[\mathcal{A}^{\rm{LR}}(-\kappa^*)]_{nn}\nonumber\\
	&=-\sum_{n=1}^{n_{\rm r}}[\mathcal{A}^{\rm{LR}}(\kappa)]_{nn}
	-i{\rm tr} [\Up^{\rm{RL}}(\kappa)^\dagger \partial_{\kappa} \Up^{\rm{LR}}(\kappa)].
\label{eq:A(k)-A(-k)}
\end{align}
The similar relation holds for $\mathcal{A}^{\rm{RL}}(\kappa)$. Taking the summation of Eq.~\eqref{eq:A(k)-A(-k)} and the one for $\mathcal{A}^{\rm{RL}}(\kappa)$, we obtain
\begin{align}
	&\sum_{n=1}^{n_{\rm r}}[\mathcal{A}(-\kappa)]_{nn}+\sum_{n=1}^{n_{\rm r}}[\mathcal{A}(\kappa)]_{nn}\nonumber\\
	&=-\frac{i}{2} \left\{ {\rm tr} [\Up^{\rm{RL}}(\kappa)^\dagger \partial_{\kappa} \Up^{\rm{LR}}(\kappa)]
	    +{\rm tr} [\Up^{\rm{LR}}(\kappa)^\dagger \partial_{\kappa} \Up^{\rm{RL}}(\kappa)]\right\} \nonumber\\
	&=-\frac{i}{2}\partial_\kappa\left[\log \,{\rm det} [\Up^{\rm{LR}}(\kappa)\Up^{\rm{RL}}(\kappa)]\right].
\end{align}

\section{Proof of of Eq.~\eqref{PHS_berry_int}\label{prof_phs}}
From the labeling rule for positive- and negative-index states introduced in the main text, 
the particle-hole counterpart of a positive-index state $|u_{n}^{\rm R,L}(\kappa)\rangle \,(n>0)$ can be expanded with respect to negative-index states $|u_{m}^{\rm R,L}(-\kappa^*)\rangle \, (m<0)$ as
\begin{subequations}
	\begin{align}
		\mathcal{C}\ket{u^{\rm{R}}_{n>0}(\kappa)}
		&=\sum_{m<0}[\Uc^{\rm{LR}}(-\kappa^*)]_{mn}\ket{u^{\rm{R}}_m(-\kappa^*)}
		,\\
		\mathcal{C}\ket{u^{\rm{L}}_{n>0}(\kappa)}
		&=\sum_{m<0}	[\Uc^{\rm{RL}}(-\kappa^*)]_{mn}\ket{u^{\rm{L}}_m(-\kappa^*)}
		,
	\end{align}
\end{subequations}
where
\begin{subequations}
\begin{align}
[\Uc^{\rm{LR}}(-\kappa^*)]_{mn}&:=\braket{u^{\rm{L}}_m(-\kappa^*)|\mathcal{C}|u^{\rm{R}}_n(\kappa)}, \\
[\Uc^{\rm{RL}}(-\kappa^*)]_{mn}&:=\braket{u^{\rm{R}}_m(-\kappa^*)|\mathcal{C}|u^{\rm{L}}_n(\kappa)}
\end{align}
\end{subequations}
are $\mathcal{N}\times\mathcal{N}$ matrices and satisfy $\Uc^{\rm{RL}}(\kappa)\Uc^{\rm{LR}}(\kappa)^{\dagger}=\bm{1}$.
Using the above relations, we can rewrite $[\mathcal{A}^{\rm{LR}}(-\kappa^*)]_{nn'}$ for $n,n'>0$ as
\begin{align}
	&[\mathcal{A}^{\rm{LR}}(-\kappa^*)]_{nn'}\nonumber\\
	&=-i\braket{\ul_{n}(-\kappa^*)|\partial_{\kappa}\ur_{n'}(-\kappa^*)}
	\nonumber\\
	&=-i\sum_{m,\ell<0}[\Uc^{\rm{RL}}(\kappa)]^{\rm T}_{nm}\braket{ \mathcal{C}\ul_{m}(\kappa)|\partial_{\kappa}\Big([\Uc^{\rm{LR}}(\kappa)]^*_{\ell n'}|\mathcal{C}\ur_{\ell}(\kappa)}\Big)
	\nonumber\\
	&=-i\sum_{m,\ell<0}[\Uc^{\rm{RL}}(\kappa)]^{\rm{T}}_{nm}\braket{ \mathcal{C}\ul_{m}(\kappa)|\partial_{\kappa}\mathcal{C}\ur_{\ell}(\kappa)}[\Uc^{\rm{LR}}(\kappa)]^*_{\ell n'}		
	\nonumber\\
	&\ \ \ -i\sum_{m,\ell<0}[\Uc^{\rm{RL}}(\kappa)]^{\rm{T}}_{nm}\partial_{\kappa}[\Uc^{\rm{LR}}(\kappa)]^*_{\ell n'}\delta_{m\ell}
	\nonumber\\
	&=-i\sum_{m,\ell<0}[\Uc^{\rm{RL}}(\kappa)]^{\rm{T}}_{nm}\braket{ \partial_{\kappa}\ur_{\ell}(\kappa)|\ul_{m}(\kappa)}[\Uc^{\rm{LR}}(\kappa)]^*_{\ell n'}
	\nonumber\\
	&\ \ \ -i\sum_{m<0}[\Uc^{\rm{RL}}(\kappa)]^{\rm{T}}_{nm}\partial_{\kappa}[\Uc^{\rm{LR}}(\kappa)]^*_{mn'}
	\nonumber\\
	&=\sum_{m,\ell<0}{[\Uc^{\rm{RL}}(\kappa)]^{\rm{T}}_{nm}}{[\mathcal{A}^{\rm{LR}}(\kappa)]^{*}_{m\ell}} {[\Uc^{\rm{LR}}(\kappa)]^{*}_{\ell n'}}
	\nonumber\\
	&\ \ \ -i\sum_{m<0}[\Uc^{\rm{RL}}(\kappa)]^{\rm{T}}_{nm}\partial_{\kappa}[\Uc^{\rm{LR}}(\kappa)]^*_{mn'},
\end{align}
from which we obtain
\begin{align}
	&\sum_{n>0}[\mathcal{A}^{\rm{LR}}(-\kappa^*)]^*_{nn}\nonumber\\
	&=\sum_{m<0}[\mathcal{A}^{\rm{LR}}(\kappa)]_{mm}
		+i{\rm tr}\left[\Uc^{\rm{RL}}(\kappa)^\dagger\partial_{\kappa}\Uc^{\rm{LR}}(\kappa)\right].
\label{eq:ALR_+}
\end{align}
The similar relation holds for $\mathcal{A}^{\rm{RL}}(\kappa)$. Taking into account that $A_\pm(\kappa)$ is always real, the summation of Eq.~\eqref{eq:ALR_+} and the one for $\mathcal{A}^{\rm{RL}}(\kappa)$ resluts in
\begin{align}
	&A_+(-\kappa^*) - A_-(\kappa)\nonumber\\
	&= \frac{i}{2}\left\{ \mbox{tr}[\Uc^{\rm{RL}}(\kappa)^{\dagger}\partial_{\kappa}\Uc^{\rm{LR}}(\kappa)]
	+ \mbox{tr}[\Uc^{\rm{LR}}(\kappa)^{\dagger}\partial_{\kappa}\Uc^{\rm{RL}}(\kappa)]\right\}\nonumber\\
	&= \frac{i}{2} \partial_\kappa \left[ \log {\rm det}[\Uc^{\rm{LR}}(\kappa)\Uc^{\rm{RL}}(\kappa)]\right]\nonumber\\
	&=-\partial_\kappa\theta_\mathcal{C}(\kappa),
\end{align}
where $\theta_\mathcal{C}(\kappa):={\rm arg}[ {\rm det}\Uc^{\rm RL}(\kappa)]$ and we have used the relation $[\Uc^{\rm{RL/LR}}(\kappa)]^{\dagger}=[\Uc^{\rm{LR/RL}}(\kappa)]^{-1}$.



\twocolumngrid
\bibliography{reference.bib}
\end{document}